\documentclass[journal]{IEEEtran}

\usepackage{amsmath}
\usepackage{url}
\usepackage{algpseudocode}
\usepackage{algorithm}
\usepackage{balance}
\usepackage{pifont}
\usepackage{amssymb}
\usepackage{subfigure}
\usepackage{multirow}
\usepackage{cite}
\usepackage[table]{xcolor}
\usepackage[justification=justified,font=small,labelfont=bf]{caption}
\usepackage{enumitem}
\usepackage{tikz,pgf}
\usepackage[export]{adjustbox}
\usetikzlibrary{calc}

\newcommand{\ignore}[1]{}

\newcommand{\blueHL}[1]{{\textcolor{blue}{#1}}}

\colorlet{LightRed}{white!70!red}
\colorlet{LightGreen}{white!70!green}

\begin{document}
\title{Performance Analysis of DNN Inference/Training with Convolution and non-Convolution Operations}

\author{Hadi Esmaeilzadeh, Soroush Ghodrati, Andrew B. Kahng, Sean Kinzer,\\ Susmita Dey Manasi$^*$, Sachin S. Sapatnekar, and Zhiang Wang
	\thanks{$^*$Primary author}
	\thanks{S. D. Manasi and S. S. Sapatnekar are with the Department of Electrical and Computer Engineering, University of Minnesota Twin Cities, Minneapolis, MN, USA. H. Esmaeilzadeh, S. Ghodrati, A. B. Kahng, S. Kinzer and Z. Wang are with the Department of Computer Science and Engineering, University of California San Diego, La Jolla, CA, USA.}
}

\maketitle

\noindent
\begin{abstract}
Today's performance analysis frameworks for deep learning accelerators suffer from two significant limitations.  First, although modern convolutional neural network (CNNs) consist of many types of layers other than convolution, especially during training, these frameworks largely focus on convolution layers only.  Second, these frameworks are generally targeted towards inference, and lack support for training operations. This work proposes a novel performance analysis framework, SimDIT, for general ASIC-based systolic hardware accelerator platforms. The modeling effort of SimDIT comprehensively covers convolution and non-convolution operations of both CNN inference and training on a highly parameterizable hardware substrate. SimDIT is integrated with a backend silicon implementation flow and provides detailed end-to-end performance statistics (i.e., data access cost, cycle counts, energy, and power) for executing CNN inference and training workloads. SimDIT-enabled performance analysis reveals 
that on a 64$\times$64 processing array, non-convolution operations constitute 59.5\% of total runtime for ResNet-50 training workload. In addition, by optimally distributing available off-chip DRAM bandwidth and on-chip SRAM resources, SimDIT achieves 18$\times$ performance improvement over a generic static resource allocation for ResNet-50 inference.
\end{abstract}	


\section{Introduction}
\label{sec:Intro}

\noindent
The success of using ASIC-based custom accelerator platforms in executing deep convolutional neural networks (DNNs/CNNs) has made them a prime choice as the computing platform for inference and training~\cite{TPUv2-2020}. The design of optimized DNN accelerators involves a large design parameter space. A comprehensive performance analysis framework is crucial to explore this space and achieve optimized hardware performance from a large spectrum of design choices. 

Modern DNN graphs, particularly for training, consist of many types of layers, e.g., convolution, non-linear activation, tensor addition for residual links, max pooling, average pooling, batch normalization, gradient computations with respect to input data and parameters, and multidimensional parameter tensor updates. While convolution layers are the largest single contributor to execution time during DNN execution, the combined contributions from layers other than convolution have non-negligible effects on the system-level performance of a network. During training, the energy and runtime overhead from the large set of non-convolution layers can even exceed the energy and runtime for convolution layers (33\%--52\% energy and 30\%--60\% runtime overhead for ResNet-50, as demonstrated later in Section~\ref{sec:ResltSim}\ignore{Table~\ref{tbl:conv_non_conv}}). Therefore, it is highly critical to develop simulation framework that models non-convolution layers along with convolution layer to enable accurate system-level performance analysis of DNN accelerators.

Prior works on DNN accelerator simulators do not comprehensively model all types of DNN layers while they target inference only and rarely have support for the training operations. For example, SCALE-Sim~\cite{ScaleSim} develops an inference simulator for ASIC-based systolic accelerators that models convolution layers only. Timeloop~\cite{Timeloop2019}, MAESTRO~\cite{Maestro2020}, and DeepOpt~\cite{DeepOptManasi} propose DNN dataflow analysis frameworks for inference accelerators, but their evaluations focus only on convolution layers, and the work in~\cite{EyerissToolPaper} proposes an energy estimation model for the convolution operation only.
While the modeling efforts in~\cite{YMa2020, AutoDNNchip2020} include pooling and tensor addition along with convolution, their scope is limited to inference and do not have support for training operations. TRIM~\cite{TRIM2022} proposes a design space explorer for DNN training, but does not support batch normalization, a heavy training workload, and is not evaluated on mainstream ASIC accelerators~\cite{Jouppi2017, TPUv2-2020, VeriGoodML, GemminiDAC2021, VTAMicro} that use systolic or vector dot-product style hardware.

\begin{table}[b]
	\centering
	\caption{Inference and training operations modeled by SimDIT.}
	\label{tbl:LayerList}
	{\scriptsize
    \resizebox{1.0\linewidth}{!}{
		\begin{tabular}{|c|c|cc|c|}
			\hline
			&
            Inference/                                                                   & \multicolumn{2}{c|}{Backward pass of training}                                                         & \multicolumn{1}{l|}{Section}                                                                             \\ 
			&
            Forward pass of training                                                     & \multicolumn{2}{c|}{}                                                                                  & \\ \hline            
			\rowcolor[HTML]{FFDBD9} 
            &
			\cellcolor[HTML]{FFDBD9}                                                      & \multicolumn{1}{c|}{\cellcolor[HTML]{FFDBD9}}                                          & ifmap        & \cellcolor[HTML]{FFDBD9}                                                                              \\ \cline{4-4}
			\rowcolor[HTML]{FFDBD9} 
            &
            \multirow{-2}{*}{\cellcolor[HTML]{FFDBD9}Convolution}                         & \multicolumn{1}{c|}{\multirow{-2}{*}{\cellcolor[HTML]{FFDBD9}Gradient of loss w.r.t.}} & weight       & \cellcolor[HTML]{FFDBD9}                                                                              \\ \cline{2-4}
            \rowcolor[HTML]{FFDBD9} 
            &
			\cellcolor[HTML]{FFDBD9}                                                      & \multicolumn{1}{c|}{\cellcolor[HTML]{FFDBD9}}                                          & ifmap        & \cellcolor[HTML]{FFDBD9}                                                                              \\ \cline{4-4}
			\rowcolor[HTML]{FFDBD9}
            \multirow{-4}{*}{\rotatebox[origin=c]{30}{Conv}} &
			\multirow{-2}{*}{\cellcolor[HTML]{FFDBD9}Fully-connected}                     & \multicolumn{1}{c|}{\multirow{-2}{*}{\cellcolor[HTML]{FFDBD9}Gradient of loss w.r.t.}} & weight       & \multirow{-4}{*}{\cellcolor[HTML]{FFDBD9}\begin{tabular}[c]{@{}c@{}}IV-C,\\ IV-D,\\ V-B\end{tabular}} \\ \hline
			\rowcolor[HTML]{D2EAD2} 
            &
			Tensor-add                                                                    & \multicolumn{2}{c|}{\cellcolor[HTML]{D2EAD2}Gradient of loss w.r.t. input}                            & \cellcolor[HTML]{D2EAD2}                                                                              \\ \cline{2-4}
			\rowcolor[HTML]{D2EAD2} 
            &
			ReLU activation                                                               & \multicolumn{2}{c|}{\cellcolor[HTML]{D2EAD2}Gradient of loss w.r.t. input}                            & \cellcolor[HTML]{D2EAD2}                                                                              \\ \cline{2-4}
			\rowcolor[HTML]{D2EAD2}
            \multirow{-3}{*}{\rotatebox[origin=c]{30}{non-Conv}}             &
			Pool (max, avg, global avg)                                           & \multicolumn{2}{c|}{\cellcolor[HTML]{D2EAD2}Gradient of loss w.r.t. input}                            & \multirow{-3}{*}{\cellcolor[HTML]{D2EAD2}IV-E}                                                        \\ \hline
			\rowcolor[HTML]{F7F7C4} 
            &
			\cellcolor[HTML]{F7F7C4}                                                      & \multicolumn{1}{c|}{\cellcolor[HTML]{F7F7C4}}                                          & input        & \cellcolor[HTML]{F7F7C4}                                                                              \\ \cline{4-4}
			\rowcolor[HTML]{F7F7C4} 
            &
			\multirow{-2}{*}{\cellcolor[HTML]{F7F7C4}Batch normalization} & \multicolumn{1}{c|}{\multirow{-2}{*}{\cellcolor[HTML]{F7F7C4}Gradient of loss w.r.t.}} & scale, shift & \cellcolor[HTML]{F7F7C4}                                                                              \\ \cline{2-4}
			\rowcolor[HTML]{F7F7C4} 
            \multirow{-3}{*}{\rotatebox[origin=c]{30}{non-Conv$^*$}}             &
			\multicolumn{1}{c|}{\cellcolor[HTML]{F7F7C4}--}                              & \multicolumn{2}{c|}{\cellcolor[HTML]{F7F7C4}Update of 1D, 2D, 4D parameters}                          & \multirow{-3}{*}{\cellcolor[HTML]{F7F7C4}V-C}                                                         \\ \hline
            \multicolumn{5}{l}{\textit{\hspace{4mm}*Training only}} 
		\end{tabular}
	}
	}
\end{table}

Furthermore, prior convolution focused modeling efforts often suffer from inaccuracy and are inadequate for handling convolution operations during the training phase (discussed in Section~\ref{sec:LossGradConv}). For example, \cite{ScaleSim, Maestro2020} do not model the stall cycles due to off-chip communication, while~\cite{Timeloop2019} adopts a very simplified modeling approach for the DRAM stalls. This can lead to inaccurate estimation of runtime, especially during training where the operations are extremely memory-intensive. 
The works in~\cite{YMa2020, DeepOptManasi} do not support tiling across the kernel height-width dimensions. While this is acceptable for the inference phase, where kernel height$\times$width is quite small (i.e., lies between 1$\times$1 to 11$\times$11 for standard CNNs), it is not sufficient to handle the convolution operation during training where kernel height$\times$width can be very large (i.e., up to 223$\times$223 during gradient computation for ResNet-50). Therefore, tiling across these kernel dimensions is necessary since they can be too large to fit in the on-chip memory.

This work proposes SimDIT, a comprehensive simulation framework for CNN inference and training for ASIC-based systolic DNN accelerator platforms. SimDIT comprehensively models convolution operations, including the attributes required to handle training, as well as a diverse set of non-convolution operations, thus enabling performance analysis and facilitating design space exploration of both convolution and non-convolution workloads of a network. We develop a generic tile-based template (discussed in Section~\ref{sec:TileTemp}) to abstract the computation of the layers. Our generic tile-based modeling strategy in SimDIT enables us to develop a tractable modeling framework that covers a large set of operations. A list of the operations/layers that are modeled by SimDIT to support CNN inference and training is summarized in Table~\ref{tbl:LayerList}. The attributes of the layers are discussed in Sections~\ref{sec:DNNinf} and \ref{sec:DNNtrain}. The source-code of SimDIT is publicly available~\cite{SimDT-sourcecode}.

\section{General Framework of SimDIT}
\label{sec:framewrk}

\begin{figure}[t]
	\centering
	\includegraphics[width=3.6in]{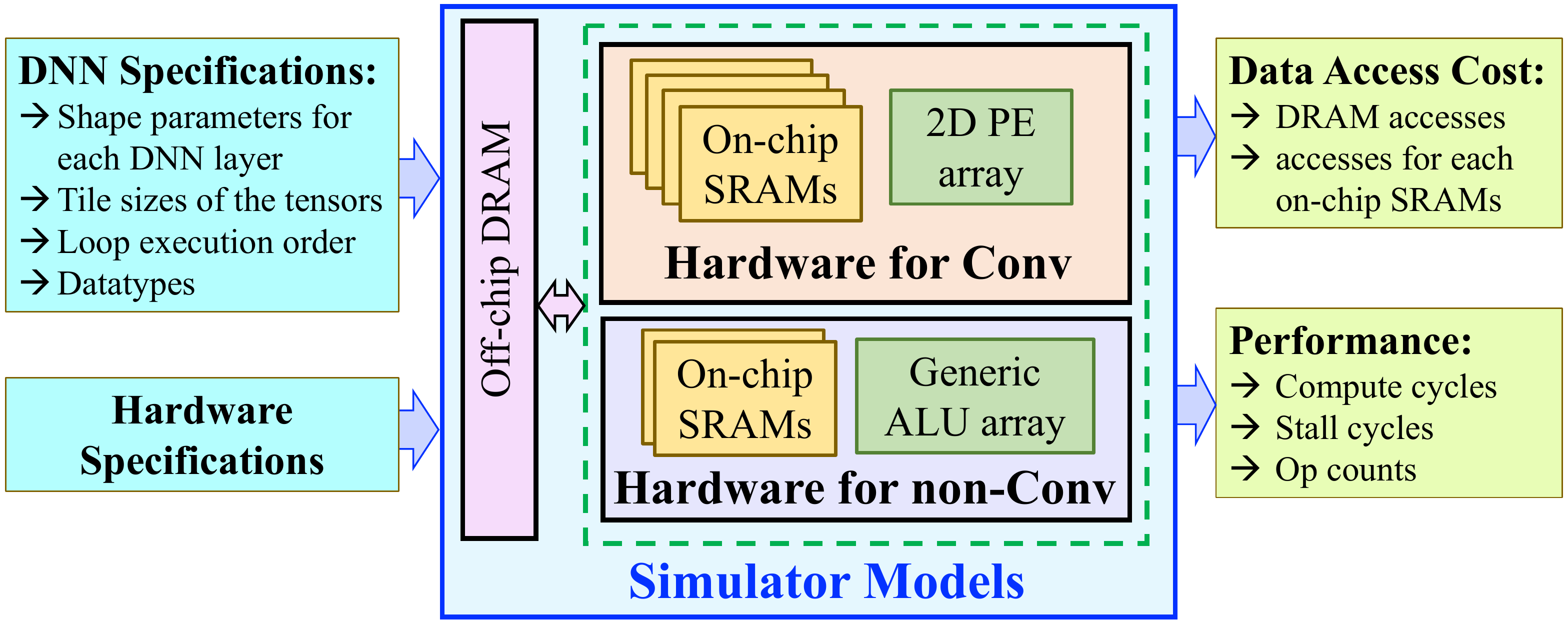}
	\vspace{-3mm}
	\caption{General overview of SimDIT.}
	\label{fig:SimDITOver}
\end{figure}

\noindent
Fig.~\ref{fig:SimDITOver} illustrates the general framework of SimDIT, our simulator for DNN inference and training
for ASIC-based DNN accelerators. ASIC accelerator platforms comprise hardware components that execute operations within a DNN workload~\cite{TPUv2-2020, VeriGoodML, IntelVTA2021}. We classify these operations as:\\
(1) \textit{Conv}, in convolution and fully-connected (FC) layers.\\
(2) \textit{non-Conv}, in layers other than convolution, such as activation (e.g., using a rectified linear unit (ReLU)), pooling (Pool), tensor addition (Tensor-Add), batch normalization (BN), gradient computation, and parameter updates.

In SimDIT, we develop models to capture the execution behavior of both Conv and non-Conv hardware components of DNN accelerators. The modeling substrate of SimDIT is fully parameterizable in terms of a range of hardware attributes, e.g, size of the compute cores, bit-widths of each type of data, on-chip buffer sizes, and DRAM bandwidth at each off-chip interface. A full list of the parameterizable hardware attributes of SimDIT is presented in Table~\ref{tbl:HrdParam} while the details about the target hardware platforms are discussed in Section~\ref{sec:HrdPlat}. 

Fig.~\ref{fig:SimDITOver} shows the high-level input and output interface of SimDIT. SimDIT takes two files (in .json format) as inputs: 
\begin{enumerate}
	\item a {\em Hardware Specifications} file that contains the specifications of the hardware attributes, and
	\item a {\em DNN Specifications} file containing description for each layer of a DNN that includes information such as dimensions of all the input/output tensors, tile sizes of the tensors, order of execution loops, and datatypes.
\end{enumerate}

\noindent
Given the above specifications, SimDIT produces end-to-end {\em performance statistics} for executing the DNN on the hardware:
\begin{itemize}
\item {\bf cycle counts:} number of on-chip cycles to perform the computations, DRAM stall cycles when the compute cores wait for data to be fetched from the off-chip memory, and total number of cycles to execute the network; 
\item {\bf on-chip access counts:} access counts for each on-chip SRAM for each type of data;  
\item {\bf off-chip access counts:} access counts for the DRAM for each type of data; and 
\item {\bf op counts:} counts for each type of arithmetic operation. 
\end{itemize}
The framework of SimDIT also includes energy and power computation module (discussed in Section~\ref{sec:EnPow}) that combines backend data from post synthesized placed-and-routed designs with the {\em performance statistics} to compute end-to-end execution energy, power, and runtime of a DNN.

\section{ASIC Hardware Platform}
\label{sec:HrdPlat}

\begin{figure}[t]
	\centering
	\includegraphics[width=2.4in]{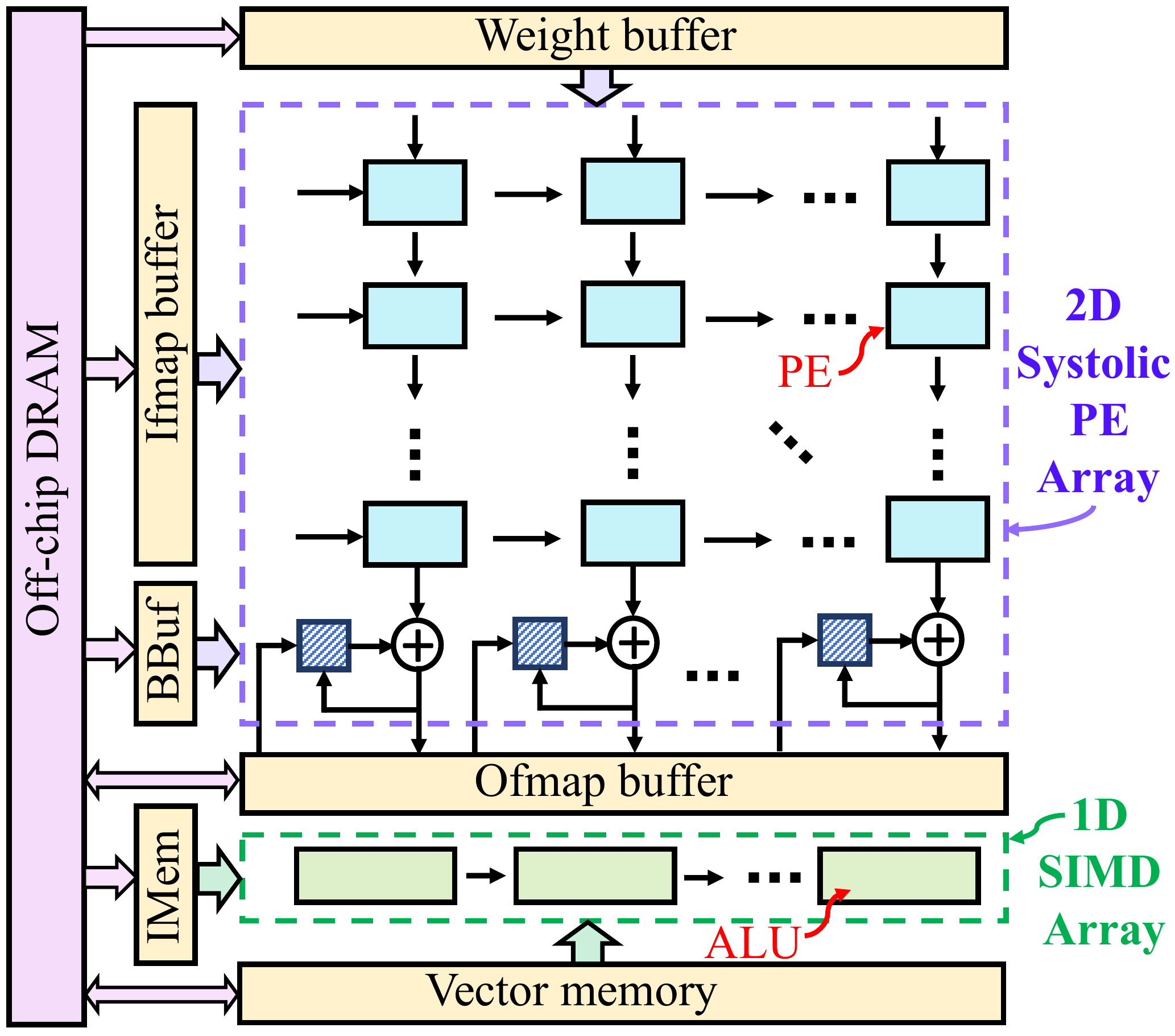}
	\vspace{-0mm}
	\caption{Block diagram of the system-level hardware architecture.}
	\label{fig:HrdArch}
	\vspace{-0mm}
\end{figure}

\noindent
Our target hardware platform is a general ASIC-based systolic DNN accelerator, similar to TPUv2~\cite{TPUv2-2020} and GeneSys~\cite{VeriGoodML}. Fig.~\ref{fig:HrdArch} shows a system-level overview of the hardware architecture. The architecture consists of two key components: a systolic array for Conv and a SIMD array for non-Conv workloads. The SIMD array here is more versatile than in inference accelerators~\cite{GemminiDAC2021, VTAMicro} since it supports a diverse set of non-Conv operations that can execute DNN training.

The systolic array executes the convolution and FC layers. The layers comprise multiple types of data: ifmap, psum/ofmap, weight, and bias that are defined in Section~\ref{sec:DNNinf}. The systolic core consists of a $J$$\times$$K$ array of processing elements (PE). Each PE contains a MAC unit and pipeline registers to forward ifmap (horizontally) and psum (vertically) data to the adjacent PEs while the weight data is held locally within each PE. In each clock cycle, the PE array performs MAC operations between a 1$\times$$J$ vector of ifmap and a $J$$\times$$K$ matrix of weight. During systolic execution, data in the ifmap vector flow horizontally (left to right) across columns and psums get accumulated vertically (top to bottom) across rows. After the pipeline is filled, the array produces a 1$\times$$K$ vector of psum/ofmap every cycle\ignore{ (i.e., one psum from every column)}. The systolic array has four on-chip buffers: IBuf, OBuf, WBuf, and BBuf for ifmap, ofmap/psum, weight, and bias data, respectively.

\begin{table}[t]
	\centering
	\caption{Hardware configuration parameters.}
	\label{tbl:HrdParam}
	\vspace{-0mm}
	{\scriptsize
		\resizebox{1.0\linewidth}{!}{
		\begin{tabular}{|l|l|l|l|}
			\hline
			Systolic array attributes             & Notations   & SIMD attributes          & Notations             \\ \hline
			Number of PE rows, columns      & $J$, $K$       & Number of ALUs                 & $K$                    \\ \hline
			\begin{tabular}[c]{@{}l@{}}Sizes of weight buffer,\\ bias buffer\end{tabular}           & $Wbuf$, $Bbuf$ & \begin{tabular}[c]{@{}l@{}}Size of vector\\ memory\end{tabular}                             & $Vmem$                 \\ \hline
			\begin{tabular}[c]{@{}l@{}}Sizes of ifmap buffer,\\ ofmap buffer\end{tabular}           & $Ibuf$, $Obuf$ & \begin{tabular}[c]{@{}l@{}}Size of instruction\\ memory\end{tabular}                        & $Imem$               \\ \hline
			Bit-width of weight, bias                            & $b_w$, $b_b$     & Bit-width of input          & $b_{in}$                  \\ \hline
			\begin{tabular}[c]{@{}l@{}}Bit-width of ifmap,\\ psum/ofmap\end{tabular}           & $b_i$, $b_p$ & Bit-width of output     & $b_{out}$                 \\ \hline
			\begin{tabular}[c]{@{}l@{}}DRAM bandwidth for\\ weight and bias buffer\end{tabular}     & $BW_w$        & \multirow{2}{*}{\begin{tabular}[c]{@{}l@{}}DRAM bandwidth\\ for vector memory\end{tabular}} & \multirow{2}{*}{$BW_v$} \\ \cline{1-2}
			\begin{tabular}[c]{@{}l@{}}DRAM bandwidth for\\ ifmap buffer, ofmap buffer\end{tabular} & $BW_i$, $BW_o$   &       &                \\ \hline
			\multicolumn{4}{|l|}{Number of cycles required for each type of arithmetic operation*}   \\ \hline
			\multicolumn{4}{l}{\textit{*notations introduced as needed inside the text}} 
		\end{tabular}
	}
	}
\end{table}

The SIMD array is a 1$\times$$K$ ALU array that executes all non-convolution layers (ReLU, Pool, Tensor-add, BN, gradient computation, and parameter updates). The SIMD array is pipelined across each ALU: the pipeline stages are similar to a general MIPS processor where the $K$ ALUs operate in parallel under a single instruction. The SIMD core is equipped with a vector memory (VMem) to store the input and output data as well as an instruction memory (IMem) to store the instructions. All six on-chip SRAMs of the systolic and SIMD cores communicate data with off-chip DRAM. Table~\ref{tbl:HrdParam} summarizes all the parameterizable attributes of the hardware platform.

\section{Performance Models: Inference}
\label{sec:PerfModel_Inf}

\subsection{Basics of CNN Inference}
\label{sec:DNNinf}

\noindent
Neural networks have two main types of phases: {\em (i)} training (Section~\ref{sec:DNNtrain}), when a network is trained to learn the model parameters, and {\em (ii)} inference, when a network is deployed for the target applications. Inference operations in a CNN primarily consists of Conv (convolution and FC) operations, nonlinear activation (e.g., ReLU), Tensor-add, and Pool layers.

\begin{figure}[t]
	\centering
	\includegraphics[width=3.0in]{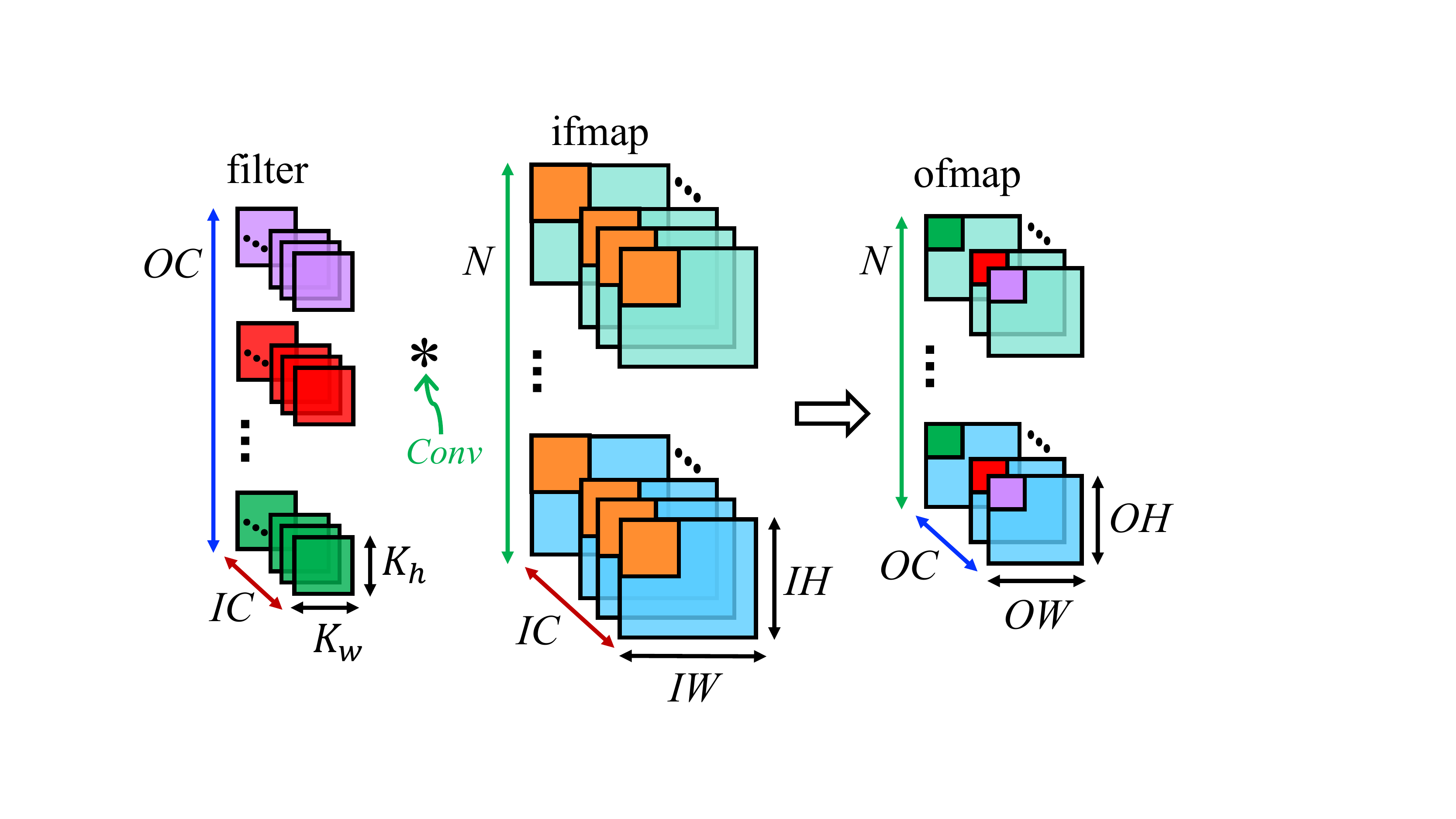}
	\vspace{-0mm}
	\caption{Convolution layer illustrating filter, ifmap, and ofmap.}
	\label{fig:ConvLr}
\end{figure}

Fig.~\ref{fig:ConvLr} illustrates the convolution layer where a {\bf filter} (or {\bf weight}) tensor of size ($K_h \times K_w \times IC \times OC$) is convolved with an input feature map ({\bf ifmap}) of size ($IH \times IW \times IC$). The computation produces an output feature map ({\bf ofmap}) of size ($OH \times OW \times OC$). During the computation, element-wise multiplication between each filter channel and a same-sized subregion of the corresponding ifmap channel (i.e., the orange ifmap region) produces intermediate partial sums ({\bf psums}) that are accumulated across all $IC$ channels to build elements in an ofmap channel (i.e., the purple ofmap element using the purple filter). Similar multiply and accumulate (MAC) operations are repeated using $OC$ 3D filters while the filter slides through the ifmap with a stride $S$ to build all the channels of a 3D ofmap. The correspondence between the filters and ofmap channels are shown by the color map. The batch size of the feature maps is shown by $N$ where similar convolution computations are concurrently performed using $N$ images. A convolution layer may include bias term of size $OC$ that gets added with each ofmap channel. An FC layer is similar to a convolution layer but has smaller tensor sizes and produces 1D ofmap.

Conv layers are optionally followed by ReLU, max/average pooling layers. A ReLU layer performs element-wise max operations on an input tensor and produces a same-sized output tensor. 
A Pool layer reduces the dimensionality of its input tensor where the computation involves max/average operations over a window of data while the window slides through each channel of the input tensor. CNNs that support residual learning (i.e., ResNet) contain shortcut connections that are realized by Tensor-add layers. The computation in a Tensor-add layer involves channel-by-channel element-wise addition, where the input and output tensors have same dimensions.

\subsection{Generic Tile-based Template}
\label{sec:TileTemp}

\noindent
We develop analytical models to obtain the end-to-end performance statistics for each CNN inference and training operation. Due to limited on-chip storage and parallel compute units (i.e., systolic PE array or SIMD ALUs) in a hardware, the computation of a layer is executed by dividing it into smaller subcomponents. We formulate a generic tile-based template that partitions the computation of each layer into {\em outer tiles} \ignore{(corresponding to the sizes of the on-chip SRAMs)} and {\em inner tiles}\ignore{ (corresponding to the number of parallel compute units)}. For each tensor of a layer, inner tile computations are repeated to cover the outer tile while outer tile computations are repeated to process the entire layer. The outer tile has the volume of data that fits in the respective on-chip SRAM, while the inner tile has the volume of data processed by the compute units in one cycle. 

\begin{figure}[t]
	\centering
	\includegraphics[width=1.00\linewidth]{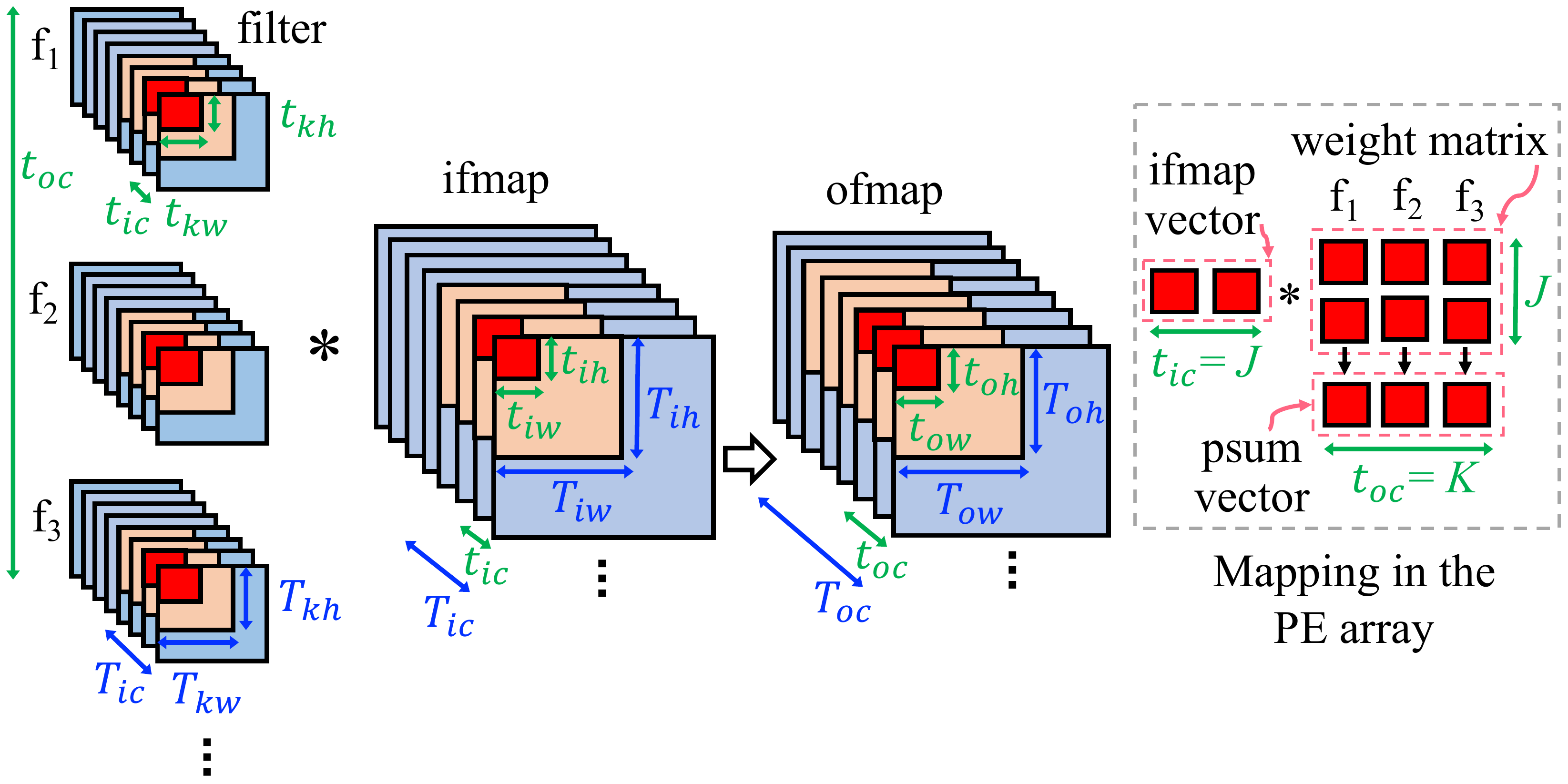}
	\caption{A tiling template for a convolution layer. The tiling parameters for batch dimension, $T_n$ and $t_n$, are omitted for simplicity.}
	\label{fig:ConvTile}
\end{figure}

Fig.~\ref{fig:ConvTile} illustrates the concept of outer and inner tiles by showing the tiling template for a convolution layer. For each type of data, light blue represents the full tensor volume, the orange region (nested within blue) illustrates the outer tile, and the red region (nested within orange) illustrates the inner tile. For each tensor, $T_{\phi}$ and $t_{\phi}$ denote the outer and inner tile sizes, respectively, for the dimensions $\phi$ where $\phi \in \lbrace oh, ow, ih, iw, n, kh, kw, ic, oc \rbrace$. Note that, our notational convention uses lower-case to indicate the {\em name of a dimension} while upper-case to denote the {\em size of a tensor along that dimension}. For example, $oh$ is the name of the height dimension of ofmap while $OH$ stands for the size of ofmap tensor along the height dimension.

\ignore{
The tiling template illustration in Fig.~\ref{fig:ConvTile} uses $T_{kh} = K_h$ and $T_{kw} = K_w$ which is the case for most Conv layers during inference since the kernel sizes are typically small (i.e., 1$\times$1, 3$\times$3). However, due to tensor transformation (discussed in Section~\ref{sec:LossGradConv}), the kernel size can be quite large during Conv training operations (e.g., 223$\times$223 for the first Conv layer of ResNet-50) where the kernel height and width dimensions may need to be split into smaller outer tiles depending on the size of the weight buffer.
}

DNN accelerators execute the convolution computation as a series of general matrix-vector multiplication (GEMM) operations~\cite{IntelVTA2021, VeriGoodML}. The right box of Fig.~\ref{fig:ConvTile} shows how the computation in the inner tile is mapped as a GEMM operation to the systolic PE array of Fig.~\ref{fig:HrdArch} to be processed in one cycle. The GEMM mapping of the convolution computation in the inner tile is performed by setting $t_{ic} = J$, $t_{oc} = K$, and all other inner tile parameters as 1 (i.e., $t_{\phi} =$ 1 where $\phi \in \lbrace oh, ow, ih, iw, n, kh, kw\rbrace$). As shown in the right box, the 1$\times J$ ifmap vector is formed using one ifmap element from each of the $J$ channels (i.e., the red region of the ifmap tensor). The $J \times K$ weight matrix is formed using $K$ 3D filters where each column of the matrix consists of data from $J$ channels of a single 3D filter (i.e., the red region of the filter tensors). Layers with small number of channels (i.e., $IC<J$, $OC<K$) are handled by zero padding the respective dimensions.
The PE array performs the computation between the ifmap vector and weight matrix and produces $K$ psums in one cycle (i.e., the red ofmap tensor region). A similar matrix-vector multiplication operation is repeated every cycle to process the outer tile. Once an outer tile is processed, the computed ofmap is moved to the DRAM and new outer tiles are loaded to the SRAMs. The process gets repeated until the entire layer is covered to produce the full ofmap volume.

We adopt similar tile-based templates to abstract the computation of each type of layer. Our generic tile-based template allows us to abstract the computation of all types of layers (as listed in Table~\ref{tbl:LayerList}), thus enabling a tractable modeling framework that supports a large number of layers/operations.

\subsection{Data Access Models for Conv: A Systolic Array Operation}
\label{sec:DataModelConv}

\noindent
{\bf DRAM access model:}
We first present our models to compute the data access cost for Conv layer, a layer executed in the systolic PE array (note that the Conv models presented in Sections~\ref{sec:DataModelConv} and \ref{sec:CycleModelConv} are used to handle Conv operations during the training phase as well, as outlined in Section~\ref{sec:LossGradConv}). We begin by computing the number of DRAM accesses for weight, ifmap, psum/ofmap, and bias data, denoted as $A_{D_w}$, $A_{D_i}$, $A_{D_p}$, and $A_{D_b}$, respectively. During the computation, outer tiles of ifmap, weight, and bias are loaded from the DRAM to their respective SRAMs. Irreducible psum tiles are stored [loaded] to [from] the DRAM as needed while the ofmaps are written back to DRAM once they are computed. The convolution computation proceeds along the seven convolution loops (i.e., $oh, ow, n, kh, kw, ic, oc$). We define seven {\em outer multipliers} to express the iterations along each loop:
\begin{align}
m_{oh} =& \; \frac{OH}{T_{oh}} \; \; ; \; \; 
m_{ow} = \frac{OW}{T_{ow}} \; \; ; \; \; 
m_{n} = \frac{N}{T_n}
\label{eq:OM}\\
m_{kh} =& \; \frac{K_h}{T_{kh}} \; \; ; \; \; 
m_{kw} = \frac{K_w}{T_{kw}} \; \; ; \; \; 
m_{ic} = \frac{IC}{T_{ic}} \; \; ; \; \; 
m_{oc} = \frac{OC}{T_{oc}}
\nonumber
\end{align}

\noindent
\underline{\textit{DRAM accesses for weight:}} Due to the large volume of weight data in CNNs, neural accelerators~\cite{Jouppi2017, VeriGoodML} typically maximize weight data reuse after fetching them from the costly off-chip DRAM. Similarly, we execute convolution loops in a weight-stationary order (i.e., $oh, ow, n$ are at the inner most loops) that maximizes weight data reuse. 
As shown in Fig.~\ref{fig:ConvTile}, the volume of outer tile for the weight, is given by 
\begin{equation}
V^o_{w-tile} = 
\textstyle \prod_{i \in \{kh, kw, ic, oc\}} T_i
\end{equation}
Through the sliding window operation, we reuse the weight data across the height, width, and batch dimension of the feature maps. Therefore, the load of weight tile from the DRAM is iterated along the remaining four convolution loops, and the number of iterations along the four convolution loops is given by the product of the corresponding multipliers:
\begin{equation}
\mathcal{M}^o_{w-tile} = \textstyle \prod_{i \in \{kh, kw, ic, oc\}} m_i
\label{eq:WeightReuseMul}
\end{equation}
The number of total DRAM access for weight is then:
\begin{equation}
A_{D_w} = V^o_{w-tile} \times \mathcal{M}^o_{w-tile} \times b_w
\label{eq:DAweight}
\end{equation}
Notice that Equation~\eqref{eq:DAweight} leads to maximal weight reuse, where each weight data is loaded only once from the DRAM. 

\noindent
\underline{\textit{DRAM accesses for ifmap:}} 
Similar to the case above, we can write the volume of the outer tile for ifmap data as:
\begin{equation}
V^o_{i-tile} = \prod_{i \in \{ih, iw, n, ic\}} T_i \\
\end{equation}
Since the weight data remains stationary, the load of ifmap tile from DRAM is repeated along all seven convolution loops as the computation proceeds along each loop where the number of iterations for ifmap data is given by the multiplier\footnote{{\scriptsize SimDIT applies the ceiling operator on appropriate outer/inner multipliers to handle corner iteration cases for each data type. For readability, we omit the ceiling operators.}}:
\begin{equation}
\mathcal{M}^o_{i-tile} = \prod_{i \in \{oh, ow, n, kh, kw, ic, oc \}} m_i
\label{eq:DIT_ifmap}
\end{equation}
\noindent
Thus, the total number of ifmap access from DRAM is:
\begin{equation}
A_{D_i} = V^o_{i-tile} \times \mathcal{M}^o_{i-tile} \times b_i
\label{eq:DAifmap}
\end{equation}
\noindent
\underline{\textit{DRAM accesses for psum:}} Outer tile psum data volume is:
\begin{align}
V^o_{p-tile} &= \prod_{i \in \{oh, ow, n, oc\}} T_i 
\end{align}
The psums are accumulated when the computation proceeds along the $kh$, $kw$, and $ic$ dimensions, and the number of iterations corresponds to the product of the corresponding multipliers. To accumulate the psums generated from the current tile with the those from previous tiles, two operations (load and store) are required for each iteration, except for the first iteration that requires only one (store) operation. The multiplier for the access count over all outer tiles is thus:
\begin{equation}
\mathcal{M}^o_{p-tile} = \left ( \prod_{i \in \{oh, ow,n,oc\}} m_i \right ) \left ( \left[ \prod_{i \in \{kh, kw, ic \}} 2 m_i \right] - 1 \right)
\label{eq:DIT_psum_1}
\end{equation}
\noindent
As before, we write the total psum DRAM accesses as:
\begin{equation}
A_{D_p} = V^o_{p-tile} \times \mathcal{M}^o_{p-tile} \times b_p
\label{eq:DApsum}
\end{equation}
\noindent
\underline{\textit{DRAM access for bias:}} 
To maximize bias data reuse, we execute convolution keeping $oc$ at the outermost loop. After loading an outer tile of bias, $T_{oc}$, from the DRAM, this allows the addition of a bias element with all the ofmap elements in a 2D channel.
Thus, each bias term is loaded once from the DRAM, and the number of DRAM accesses for bias data is:
\begin{equation}
A_{D_b} = T_{oc} \times m_{oc} \times b_b
\end{equation}

\noindent
{\bf SRAM access model:} We now present our model to compute the number of SRAM accesses for weight, ifmap, psum/ofmap, and bias data, denoted as  $A_{S_w}$, $A_{S_i}$, $A_{S_p}$, and $A_{S_b}$, respectively. In each clock cycle, the PE array reads inner tiles of weight and ifmap data from their respective SRAMs to perform the matrix-vector multiplication operation. The PE array also performs read and write access of inner tiles of psums from the ofmap buffer as needed. Similar to Equations~\eqref{eq:OM}, we define {\em inner multipliers} to express the iterations along the seven convolution loops to cover an outer tile as follows:
\begin{align}
r_{oh} =& \; \frac{T_{oh}}{t_{oh}} \; \; ; \; \; 
r_{ow} = \frac{T_{ow}}{t_{ow}} \; \; ; \; \; 
r_{n} = \frac{T_n}{t_n}
\label{eq:InM} \\
r_{kh} =& \; \frac{T_{kh}}{t_{kh}} \; \; ; \; \; 
r_{kw} = \frac{T_{kw}}{t_{kw}} \; \; ; \; \; 
r_{ic} = \frac{T_{ic}}{t_{ic}} \; \; ; \; \; 
r_{oc} = \frac{T_{oc}}{t_{oc}}
\nonumber
\end{align}
\noindent
Furthermore, the number of inner iterations along the seven convolution loops to process a single outer tile of ofmap is computed by the multiplier, $\mathcal{M}_{inner}$, while the number of such outer iterations to cover the entire ofmap tensor is determined by the multiplier, $\mathcal{M}_{outer}$, as follows:
\begin{align}
\mathcal{M}_{inner} &= \prod_{i \in \{ oh, ow, n, kh, kw, ic, oc\}} r_i
\label{eq:M_inner}\\
\mathcal{M}_{outer} &= \mathcal{M}^o_{i-tile}
\label{eq:M_outer}
\end{align}
\noindent
Since the PE array accesses the SRAMs during the computation of each inner tile that is nested within the outer tile, $\mathcal{M}_{inner}$ along with $\mathcal{M}_{outer}$ is needed to determine the number of SRAM accesses. In Table~\ref{tbl:SRAMAccessConv}, we summarize the computation of SRAM accesses associated with each type of data. The first column shows the volume of the inner tile, $V^i_{\alpha-tile}$, for each data type where $\alpha \in \lbrace w, i, p, b \rbrace$ indicating weight, ifmap, psum, and bias, respectively. The blue terms in parentheses in the second column represent the required multiplier to obtain the respective number of SRAM accesses.

\begin{table}[tb]
	\vspace{-0mm}
	\centering
	\caption{Computation of SRAM access count for each type of data for a Conv layer.}
	\label{tbl:SRAMAccessConv}
	\vspace{-0mm}
	{\scriptsize	
		\begin{tabular}{|l|c|c|}
			\hline
			\multicolumn{1}{|c|}{Data} & \multicolumn{1}{c|}{Inner tile volume} & \multicolumn{1}{c|}{Number of SRAM accesses, $A_{S_{\alpha}}$}             \\ 
       			                     & \multicolumn{1}{c|}{$V^i_{\alpha-tile}$} &              \\ \hline
			weight                     & $\displaystyle \prod_{i \in \{kh, kw, ic, oc \}} t_i$      & $V^i_{w-tile} \cdot \blueHL{(\mathcal{M}_{inner} \cdot \mathcal{M}_{outer})} \cdot b_w$                      \\ \hline
			ifmap                      & $\displaystyle \prod_{i \in \{ih, iw, n, ic \}} t_i$       & $V^i_{i-tile} \cdot \blueHL{(\mathcal{M}_{inner} \cdot \mathcal{M}_{outer})} \cdot b_i$    \\ \hline
            psum                       & $\displaystyle \prod_{i \in \{oh, ow, n, oc \}} t_i$       & \begin{tabular}[c]{@{}l@{}}$\Big[V^i_{p-tile} \cdot \blueHL{(2 \cdot \mathcal{M}_{inner} \cdot \mathcal{M}_{outer})}$ \\ \hspace{4mm} $- (OH \cdot OW \cdot N \cdot OC)\Big] \cdot b_p$ \end{tabular} \\ \hline            
			bias                       & \multicolumn{1}{c|}{--}       & $OH \cdot OW \cdot N \cdot OC \cdot b_b$              \\ \hline
		\end{tabular}	
	}
\end{table}

For weight and ifmap, the PE array reads the inner tile volume of data every cycle, which is repeated along all seven convolution loops for both inner and outer iterations. Therefore, the multipliers for weight and ifmap account for these inner iterations nested within the outer iterations. The same process is applicable for the psum data where the PE array performs both read and write access every cycle (i.e., 2$\times$ higher access than weight or ifmap). However, for each ofmap element, only write access is required when the associated psum is generated for the first time. The subtraction of the second term in the psum SRAM access expression incorporates this behavior. During the addition of bias data, the PE array reads a bias element from BBuf for each ofmap element. Therefore, the size of ofmap tensor directly provides the expression for bias SRAM access.

\subsection{Cycle Count Models for Conv: A Systolic Array Operation}
\label{sec:CycleModelConv}

\noindent
The total time to execute a layer is the sum of two components: {\em (i)}~computation cycles and {\em (ii)}~DRAM stall cycles.

\noindent
{\bf Model for computation cycles:} The number of MAC operations to process an outer tile is given by $(\prod_{i \in \{oh,ow,n,oc\}} T_i) \times (\prod_{i \in \{kh, kw, ic\}} T_i )$,
where the term inside the second parenthesis represents the number of MAC operations associated with each ofmap location of the tile. Since the PE array performs $J \times K$ (i.e., $t_{ic} \times t_{oc}$, shown in Fig.~\ref{fig:ConvTile}) MAC operations in parallel, the number of cycles to compute an outer tile is given by~\eqref{eq:Comp_tile}, where the $ic$ and $oc$ dimensions are mapped along the rows and columns of the PE array, respectively. 
\begin{equation}
{\cal C}^{Conv}_{tile} = 
\left ( \textstyle \prod_{i \in \{oh, ow, n, kh, kw\}} T_i \right ) \times
\left \lceil \frac{T_{ic}}{J} \right \rceil \times \left \lceil \frac{T_{oc}}{K} \right \rceil
\label{eq:Comp_tile}
\end{equation}
The number of outer tiles required to process a Conv layer is given by the multiplier $\mathcal{M}_{outer}$ (shown in Equation~\eqref{eq:M_outer}). Thus, the total number of computation cycles for a layer is:
\begin{equation}
{\cal C}^{Conv} = \Big[{\cal C}^{Conv}_{tile} + PSO_{SA} \Big] \times M_{outer}
\label{eq:Comp_layer}
\end{equation}
For each outer tile, $(J-1) + (K-1)$ initial cycles are required to fill the pipeline of the vertical and horizontal systolic execution in the PE array. This pipeline setup overhead per tile in the systolic array is denoted by $PSO_{SA}$ in~\eqref{eq:Comp_layer}.

\begin{figure}[t]
	\vspace{-0mm}
	\centering
	\includegraphics[width=0.8\linewidth]{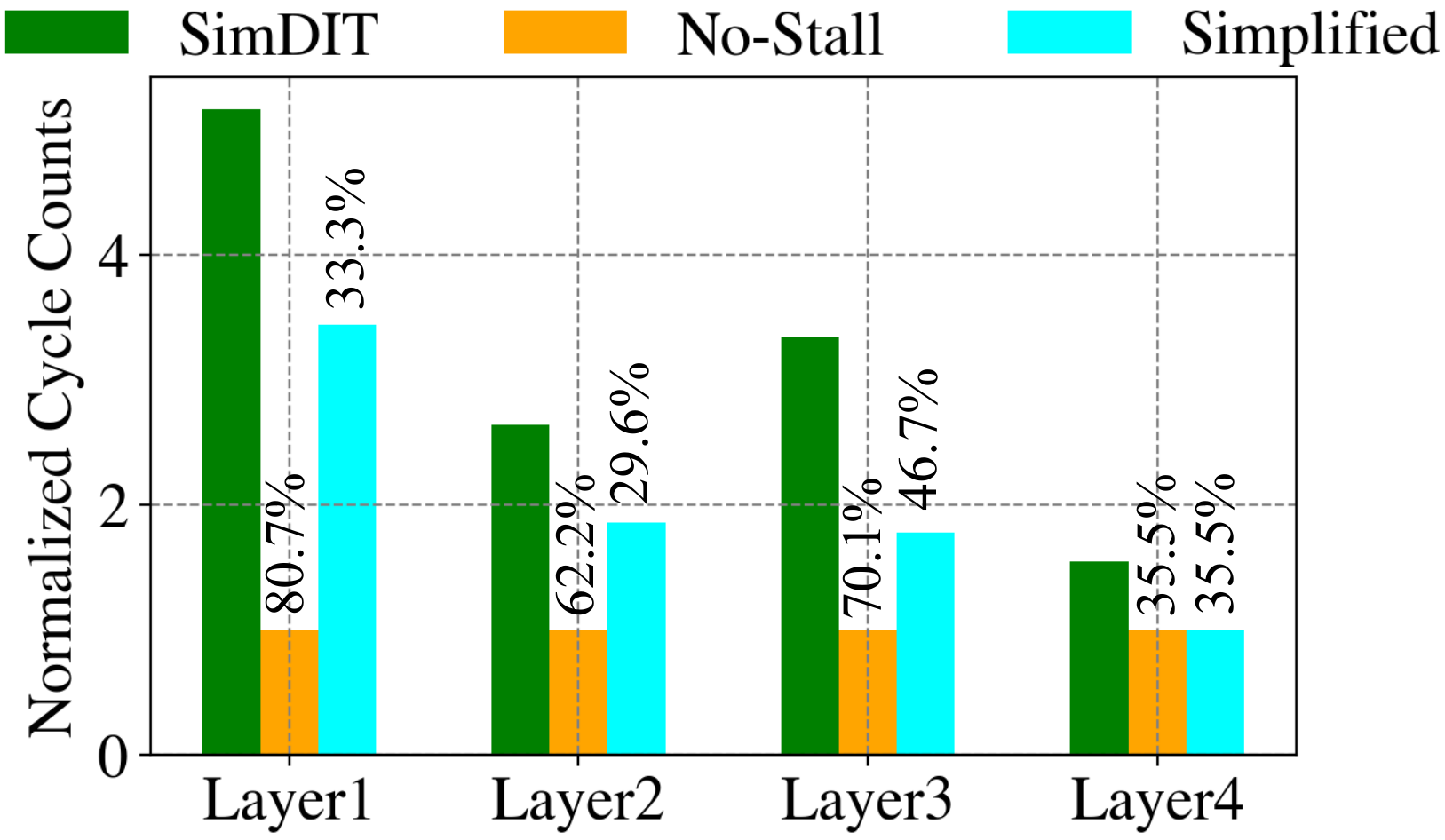}
	\vspace{-0mm}
	\caption{Comparison of SimDIT cycle counts with No-Stall and Simplified cases using four Conv layers of ResNet-50 for both inference (Layer1, Layer2) and training (Layer3, Layer3) phases.}
	\label{fig:Conv_compare}
\end{figure}

\begin{figure*}[t]
	\vspace{-0mm}
	\centering
	\includegraphics[width=4.6in]{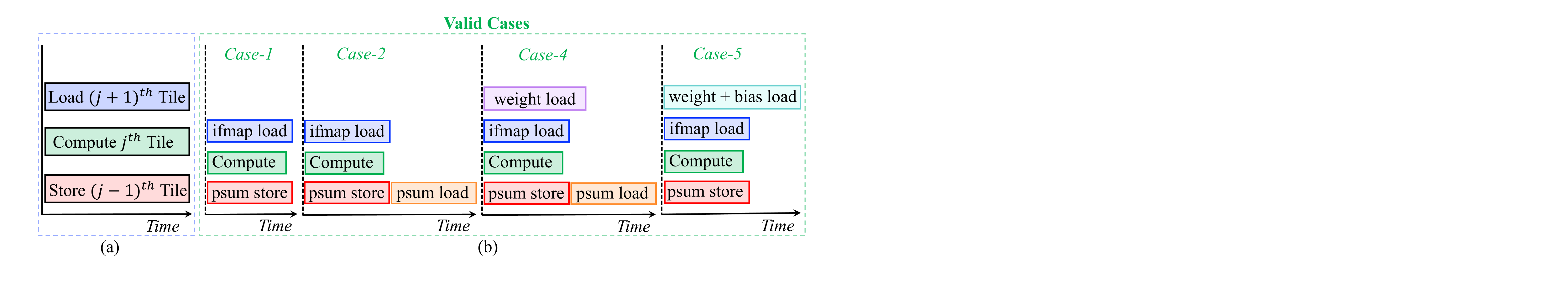}
	\vspace{-0mm}
	\caption{(a) Illustration of parallel load, compute, and store operations of tiles. (b) Four unique load-store scenarios while processing convolution tiles.}
	\label{fig:DRAMStall}
\end{figure*}

\noindent
{\bf Model for DRAM stall cycles:} Due to limited off-chip bandwidth, there are stall cycles when the PE array remains idle and waits for the data to be fetched from the DRAM. To reduce such stall cycles, all the buffers of the PE array (i.e., IBuf, OBuf, WBuf, and BBuf) are double-buffered, where the computation is overlapped with data load and store operations with the DRAM. Since DRAM stall cycles depend on the outer tiles only, in the context of DRAM stall modeling, we use to term {\em tile} to refer to the {\em outer tile}. As shown in Fig.~\ref{fig:DRAMStall}(a), under the double buffered scheme, during the computation of $j^{\rm th}$ tile, the required data to compute the next tile is loaded while the generated data from the previous tile is stored. 

In a Conv layer, data load/store patterns from/to DRAM are different during the execution of various tiles. After loading from DRAM, a data tile can be reused across multiple subsequent tile computing segments, resulting in nonuniformity in the load-store pattern across tiles. For example, for a tile that requires the weights to be loaded, the memory operations during the computation are the weight load, ifmap load, psum load, and psum store. On the other hand, for another tile that reuses weights from the previous tile, the ifmap load, psum load, and psum store take place without a weight load operation. Due to these differences in the DRAM access pattern across different tiles, it is important to handle each {\em tile computing segment} with a {\em unique load-store pattern} separately to accurately estimate the DRAM stall cycles.

In Fig.~\ref{fig:Conv_compare}, we illustrate the importance  of a performance model that incorporates the hardware execution behavior at a tile-level granularity using quantitative examples. The figure compares SimDIT (a tile-level granular model) with two scenarios: {\em (i)} No-Stall: assumes zero DRAM stall cycles due to double buffering, {\em (ii)} Simplified: estimates overall cycle count using maximum of isolated execution cycles across total load/store and compute operations (i.e., maximum of four parallel components: total MAC computation cycles, total cycles for weight+bias access, total cycles for ifmap access, and total cycles for psum/ofmap access). 
	
The cycles counts (normalized to No-stall case) are shown using four representative Conv layers of ResNet-50: Layer1 and Layer2 are from inference phase while Layer3 and Layer4 are from training phase (i.e., during backward pass). The layers are executed on inference and training hardware, respectively, with 64$\times$64 PE array. The data is obtained using various breakdown of the performance statistics produced by SimDIT (i.e., computation cycles, DRAM stalls, and total off-chip access counts for each type of data).
As can be seen from Fig.~\ref{fig:Conv_compare}, No-stall and Simplified cases underestimate the cycle counts to a large degree leading to inaccurate performance estimation: up to 80.7\% for No-stall while up to 46.7\% for the Simplified case. Therefore, it is essential to adopt tile-level granular model that comprehensively captures the behavior of each tile computing segment.

\begin{table}[htb]
	\vspace{-0mm}
	\centering
	\caption{An exhaustive enumeration of three types of load/store operations. The last column marks the valid/invalid combinations.}
	\label{tbl:DRAMStallCases}
	\vspace{-0mm}
	\begin{tabular}{|c|c|c|c|c|}
		\hline
		Cases  & {\em weight+bias load} & {\em weight load} & {\em psum load} & Validity \\ \hline
		Case-1 & 0                & 0           & 0          & $\checkmark$        \\ \hline
		Case-2 & 0                & 0           & 1          & $\checkmark$        \\ \hline
		Case-3 & 0                & 1           & 0          & $\times$        \\ \hline
		Case-4 & 0                & 1           & 1          & $\checkmark$        \\ \hline
		Case-5 & 1                & 0           & 0          & $\checkmark$        \\ \hline
		Case-6 & 1                & 0           & 1          & $\times$        \\ \hline
		Case-7 & 1                & 1           & 0          & $\times$        \\ \hline
		Case-8 & 1                & 1           & 1          & $\times$        \\ \hline
	\end{tabular}
\end{table}

In our weight-stationary execution of convolution loops, since the weight data remains stationary, movement of the other two types of data, {\em ifmap load} and {\em psum store}, takes place during the execution of all tiles. There are three remaining load/store operations, {\em weight+bias load} (load operation of weight and bias tile both), {\em weight load} (load operation of weight tile only without the load operation of bias tile)\footnote{{\scriptsize In standard CNNs, {\em bias load} is a small fraction of {\em weight load} operation. For example, for ResNet-50 inference on a 64$\times$64 array, \#of bias access from DRAM is two orders of magnitude lower than \#of weight access. Despite this, we treat {\em weight+bias load} and {\em weight load} as separate cases for the completeness of the model.}}, and {\em psum load} (load operation of psum tile), that may or may not occur during the computation of a tile depending on the reuse of the respective data for that tile. Therefore, tile computing segments with unique load-store patterns are determined by the combinations of these three load/store operations. Exhaustive enumeration of these 3 operations gives 8 cases with unique load-store patterns. The properties of the cases are summarized in Table~\ref{tbl:DRAMStallCases} where, for each column, bit 1 [0] indicates that the corresponding load/store operation takes place [does not take place] during a tile computing segment. Now, 4 cases out of the 8 (indicated in the last column of the table) cannot occur during the execution of a tile due to the following properties:
\begin{itemize}
	\item {\em weight+bias load} and {\em weight load} are mutually exclusive since weight load operation is already included within the operation where both weight and bias are being loaded. This makes Case-7 and Case-8 invalid.
	\item To maximize bias data reuse, the bias tile is loaded during the processing of the outermost $oc$ loop (discussed under ``DRAM access model'' in Section~\ref{sec:DataModelConv}) when the $kh, kw$, and $ic$ loops (i.e, the dimensions along which psum is accumulated) go through their first iteration. Since the first iteration along these three loops does not have any previously generated psum to load, {\em psum load} and {\em weight+bias load} operations are mutually exclusive. This makes Case-6 invalid.
	\item {\em weight load} occurs when the computation proceeds along the $kh, kw$, and $ic$ loops starting from the second iteration along these dimensions. Since for these computations, \ignore{loops, except the first iteration,} both load and store operations of psum are required to perform psum accumulation, Case-3, where {\em weight load} without an accompanying {\em psum load} occurs, is invalid.
\end{itemize}
\noindent
A representation of the remaining four valid cases illustrating the parallel computation and load/store operations is shown in Fig.~\ref{fig:DRAMStall}(b). For example, in Case-2, the {\em ifmap load}, {\em Compute}, and {\em psum store} can be carried out in parallel, but {\em psum load} must follow {\em psum store}.
For each of these four valid cases, we calculate the number of times each case occurs (i.e., occurrence count, ${\cal O}_{Case-i}$) and the associated stalls per tile. For a single case, the per tile stall cycle is multiplied by the occurrence count to obtain the total stall cycles for that case. Finally, the sum of stall cycles from all four cases provides the total DRAM stall cycles to fully process the Conv layer.

As an example, we 
illustrate the derivation of occurrence count and per tile stall cycle for one example case, Case-4.

\noindent
\underline{\textit{Number of occurrences:}} As shown in Fig.~\ref{fig:DRAMStall}(b), the unique attribute of Case-4 is the {\em weight load} operation, which only occurs in this case out of all cases in the figure. Therefore, the number of tile computing segments when {\em weight load} operation takes place provides the occurrence count for Case-4. As discussed in Section~\ref{sec:DataModelConv}, the weight is reused along $ow, oh$, and $n$ dimensions. Therefore, the number of tiles when weight data is loaded, regardless of bias load, is given by the multiplier $\mathcal{M}^o_{w-tile}$ (Equation~\eqref{eq:WeightReuseMul}).

\ignore{
\begin{equation}
\prod_{i \in \{ kh, kw, ic, oc \}} m_i
\label{eq:Multiplier1_C3}
\end{equation}
}

The bias tiles are loaded, along with the weights, with the outermost $oc$ loop. Hence, the number of tiles when {\em weight+bias load} takes place is given by $m_{oc}$. Now, by subtracting this from $\mathcal{M}^o_{w-tile}$,
we obtain the number of tiles when {\em weight load} operation occurs. Therefore, the occurrence count for Case-4 is given by:
\begin{equation}
{\cal O}_{Case-4} = \mathcal{M}^o_{w-tile} - m_{oc}
\label{eq:NT_case-3}
\end{equation}

\noindent
\underline{\textit{Number of stall cycles per tile:}} The systolic PE array in Fig.~\ref{fig:HrdArch} is equipped with separate off-chip interface for IBuf (which loads ifmap), OBuf (which loads and stores psum), and a shared interface for WBuf and BBuf (which loads weight and bias) to parallelly communicate data with the DRAM. For each case, Fig.~\ref{fig:DRAMStall}(b) pictorially shows parallel load-store operations that are overlapped with each other and with the computation. Therefore, the number of stall cycles per tile for Case-4 is computed by taking the maximum among the time required to complete the respective parallel operations and given by:
%
\begin{eqnarray}
{\cal S}^{Conv}_{tile-Case-4} =& \max \left ( {\cal C}^{Conv}_{tile}, \left \lceil \frac{V^o_{w-tile} b_w}{BW_{w}} \right \rceil, \left \lceil \frac{V^o_{i-tile} b_i}{BW_{i}} \right \rceil, \right . \nonumber \\
&\left . \left \lceil \frac{2 V^o_{p-tile} b_p}{BW_{o}} \right \rceil \right )
\label{eq:Tstall_case3}
\end{eqnarray}

\noindent
As defined in Table~\ref{tbl:HrdParam}, $BW_{w}$, $BW_{i}$, and $BW_{o}$ are the DRAM bandwidth \ignore{in bits/cycle} for weight, ifmap, and psum data, respectively. 

The total stall cycles for Case-4 are now computed by taking the product of the occurrence count in~\eqref{eq:NT_case-3} and the per-tile stall cycle count in~\eqref{eq:Tstall_case3}. Following similar procedure, we compute the stall cycles associated with the other cases.

\subsection{Models for Tensor-add (non-Conv) in the SIMD Array}
\label{sec:TensorAddModel}

\begin{figure}[t]
	\vspace{-0mm}
	\centering
	\includegraphics[width=1.00\linewidth]{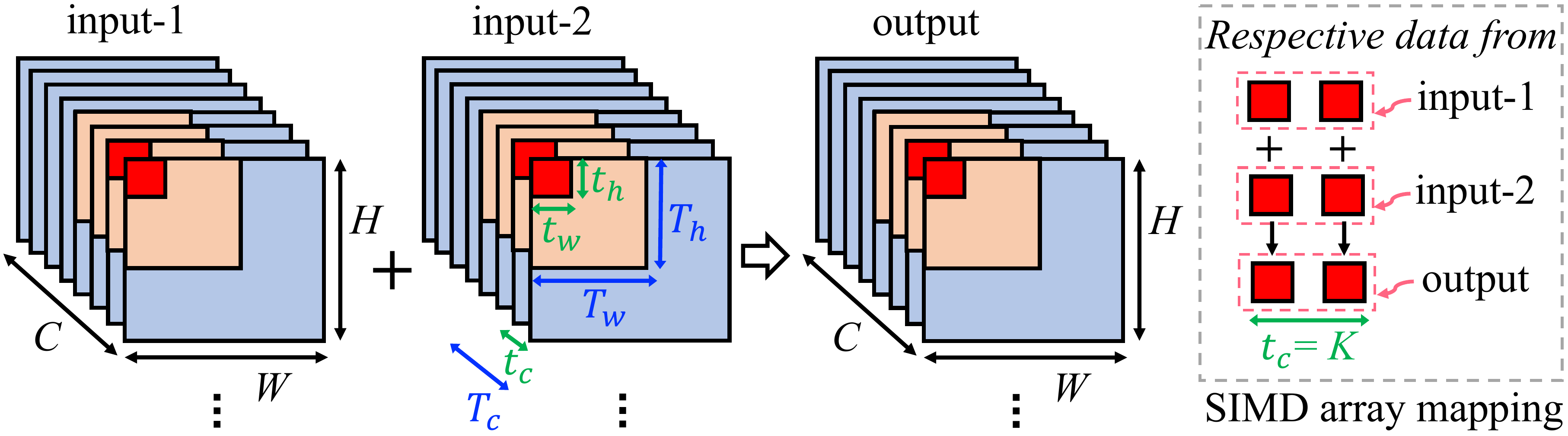}
	\vspace{-0mm}
	\caption{Tensor-add layer illustrating the two inputs, output, and their notations. The notations for the batch dimension (i.e., batch size: N; tiling parameters: $T_n$ and $t_n$) are omitted for readability.}
	\label{fig:TensorAddTile}
\end{figure}

\noindent
We now present our data access and cycle count models for the Tensor-add layer, a representative layer executed on the SIMD array component of the hardware. Similar models are developed for other non-Conv inference operations.
Fig.~\ref{fig:TensorAddTile} shows our notations for the Tensor-add layer where element-wise addition between two inputs each with size ($H \times W \times N \times C$) produces an output of the same dimension. The interpretation of the tiling template is similar to the approach discussed in Section~\ref{sec:TileTemp} where in Fig.~\ref{fig:TensorAddTile}, for each tensor, $T_{\phi}$ and $t_{\phi}$ indicate the sizes of outer and inner tile, respectively, for the dimension $\phi$ where $\phi \in \lbrace h, w, n, c\rbrace$. Since the SIMD array uses the vector memory to store all inputs and outputs of a layer, outer tiles from all inputs/outputs must fit in VMem. The right box in Fig.~\ref{fig:TensorAddTile} illustrates the mapping of the computation in the inner tile (i.e., processed in one cycle) to the $K$ ALUs of the SIMD array by setting $t_h$ = $t_w$ = $t_n$ = 1, and $t_c$ = $K$.

\noindent
\underline{\textit{Number of DRAM accesses:}} During the computation of Tensor-add layer, outer tiles of input-1 and input-2 are loaded from the DRAM to VMem while outer tiles of output are written back to DRAM as they are computed. Since the computation proceeds along the four computation loops (i.e., $h, w, n, c$), the number of outer iterations\footnote{{\scriptsize Unlike Conv layer, Tensor-add layer does not have any data reuse property and the modeling approach is less intricate than the Conv layer (this is true for other non-Conv inference layers as well). Therefore, in order to avoid unneeded notational complexity, we present the Tensor-add models using the outer tiles only, and omit the use of superscript/subscript to differentiate between inner and outer components.}} to cover the entire output tensor is given by:
\begin{equation}
{\cal M} = \frac{H}{T_h} \times \frac{W}{T_w} \times \frac{N}{T_n} \times \frac{C}{T_c}
\label{eq:SIMD_outerMul}
\end{equation}
For both the inputs as well as output, the volume of an outer tile, ${\cal V}_{tile}$, is given by ($T_h \times T_w \times T_n \times T_c$). Each tile is loaded/stored from/to DRAM once. Therefore, the number of DRAM access for the entire layer is computed by:
\begin{equation}
A_{D_{Tensor-add}} = {\cal V}_{tile} \times {\cal M} \times (2 \times b_{in} + b_{out})
\label{eq:DRAM_access_TA}
\end{equation}

\noindent
\underline{\textit{Number of SRAM accesses:}} For each outer tile of output, the number of arithmetic addition operations equals ${\cal V}_{tile}$ since there is one addition operation for each output element. Besides, for each addition operation, two input operands are read from the VMem while one output operand is written to VMem. Therefore, the number of accesses for VMem, $A_{S_{Tensor-add}}$, for the entire layer is the same as in~\eqref{eq:DRAM_access_TA}.

\noindent
\underline{\textit{Number of computation cycles:}} The SIMD array performs $K$ (i.e., $t_c$ = $K$, shown in Fig.~\ref{fig:TensorAddTile}) addition operations in parallel where $\lambda_{add}$ cycles are required to complete an addition operation per ALU. Therefore, the number of cycles to compute an outer tile is given by the following equation where the $c$ dimension is mapped along the ALUs of the SIMD array:
\begin{equation}
{\cal C}^{Tensor-add}_{tile} = (T_h \times T_w \times T_n) \times \left \lceil \frac{T_c}{K} \right \rceil \times \lambda_{add}
\label{eq:Comp_tile_TensorAdd}
\end{equation}
Since the number of outer tiles to process the entire output tensor is given by the multiplier, $\cal M$, the total number of computation cycles for the Tensor-add layer is determined by \eqref{eq:Comp_TensorAdd}. The SIMD array uses a 6-stage MIPS pipeline (i.e., instruction fetch, decode, address generation, data read, execution, and write back). In addition, the array is pipelined across the $K$ ALUs. Therefore, at the beginning of the computation of each outer tile, the array requires (6 $-$ 1) + ($K - $ 1) cycles to fill the pipeline. This pipeline setup overhead per tile, ($PSO_{SIMD}$), is incorporated in Equation~\eqref{eq:Comp_TensorAdd}.
\begin{equation}
{\cal C}^{Tensor-add} = \Big[{\cal C}^{Tensor-add}_{tile} + PSO_{SIMD} \Big] \times {\cal M}
\label{eq:Comp_TensorAdd}
\end{equation}

\noindent
\underline{\textit{Number of stall cycles:}} Since the data reuse opportunity in non-Conv layers (minimal to zero for most layers) is far less than Conv layers and VMem needs to fit tiles from all input and output tensors of a non-Conv layer, we adopt single buffering for the SIMD processor where data load/store operations with DRAM and computation take place sequentially. The SIMD array is stalled between the computation of each outer tile to load and store the respective outer tiles of data. As shown in~\eqref{eq:Stall_TensorAdd}, the associated stall cycles for the layer are computed by dividing the data volumes with the off-chip bandwidth for VMem (i.e., $BW_v$ bits/cycle).
\begin{equation}
{\cal S}^{Tensor-add} = \left \lceil \frac{{\cal V}_{tile} \times (2 \times b_{in} + b_{out})}{BW_v}\right \rceil \times {\cal M}
\label{eq:Stall_TensorAdd}
\end{equation}

\section{Performance Models: Training}
\label{sec:PerfModel_Train}

\subsection{Basics of CNN Training}
\label{sec:DNNtrain}

\noindent
Standard CNN training algorithm adopts backpropagation with stochastic gradient descent (SGD)~\cite{krizhevsky2012, ResNet}. The training method consists of many iterations of two main steps: forward pass and backward pass. The forward pass is functionally same as the inference phase where the computation advances from the first layer to the last layer of a network. The forward pass includes all the operations of the inference phase (i.e., inference is a subset of training). In addition, it also includes a BN layer that improves both training time and accuracy~\cite{BatchNorm2015}.

\begin{figure}[t]
	\vspace{-0mm}
	\centering
	\includegraphics[width=3.4in]{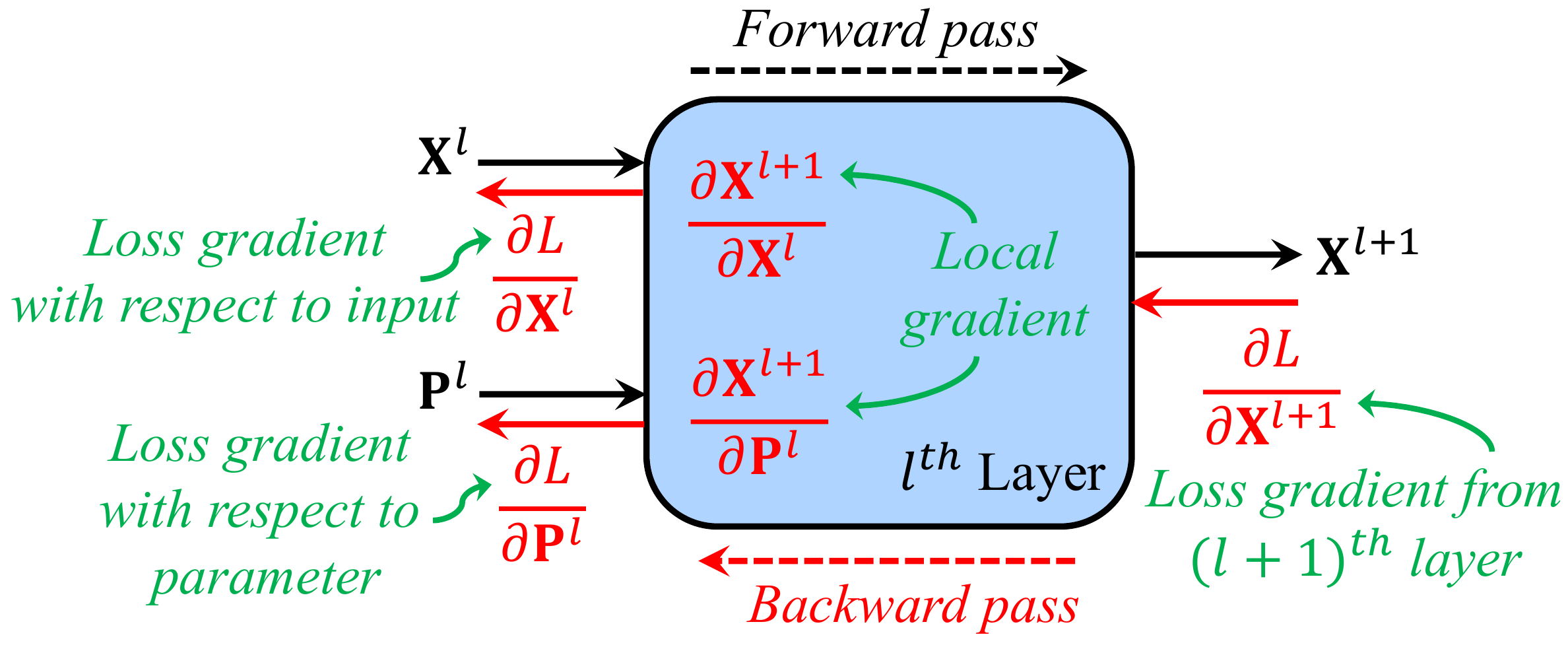}
	\vspace{-0mm}
	\caption{Overview of computation for a layer during training.}
	\label{fig:TrainLr}
\end{figure}

During the backward pass, the network is run in reverse where the computation advances from the last layer to the first layer, and a set of errors (or loss values) computed using a loss function backpropagates through each layer of the network. Fig.~\ref{fig:TrainLr} shows the computations involved for $l^{th}$ layer during training. The computation between input ${\bf X}^l$ and parameter ${\bf P}^l$ produces the output ${\bf X}^{l+1}$ during the forward pass. The backward pass performs the computation of gradient of loss (i.e., loss gradient) with respect to input ($\frac{\partial L}{\partial {\bf X}^l}$) and parameter ($\frac{\partial L}{\partial {\bf P}^l}$) using chain rule of differentiation as follows~\cite{BackProp1998}:
\begin{align}
	\frac{\partial L}{\partial {\bf X}^l} =& \; \frac{\partial L}{\partial {\bf X}^{l+1}} \cdot \frac{\partial {\bf X}^{l+1}}{\partial {\bf X}^l} \; \; ; \; \;
	\frac{\partial L}{\partial {\bf P}^l} = \; \frac{\partial L}{\partial {\bf X}^{l+1}} \cdot \frac{\partial {\bf X}^{l+1}}{\partial {\bf P}^l}
	\label{eq:chrule}
\end{align}
\noindent
Here, $L$ is the loss function, $\frac{\partial L}{\partial {\bf X}^{l+1}}$ is the loss gradient propagated from the $(l+1)^{th}$ layer while $\frac{\partial {\bf X}^{l+1}}{\partial {\bf X}^l}$ and $\frac{\partial {\bf X}^{l+1}}{\partial {\bf P}^l}$ are local gradients of the $l^{th}$ layer with respect to input and parameter, respectively. All bold notations indicate multidimensional tensors. Since $\frac{\partial L}{\partial {\bf X}^{l+1}}$ acts as an input for the $l^{th}$ layer, the gradient expressions in~\eqref{eq:chrule} are self-contained where only the local gradients need to be determined which depend on the layer type. Similar to inference, each layer can be treated individually during the backward pass without the need for the information about the loss function or next/previous layer.

The computed loss gradient with respect to parameter is used to update the parameters of the layer while the loss gradient with respect to input gets propagated to the $(l-1)^{th}$ layer. Layers that do not have any parameter (i.e., ReLU, Pool, Tensor-add) exclude the computation of $\frac{\partial L}{\partial {\bf P}^l}$. 
As stated earlier, Table~\ref{tbl:LayerList} summarizes the list of operations/layers required for CNN training along with inference.

\subsection{Loss Gradients for Conv: Systolic Array Operations}
\label{sec:LossGradConv}

\noindent
The backward pass through the $l^{th}$ Conv layer consists of two components: {\em (i)} computation of loss gradient with respect to ifmap tensor ${\bf X}^{l}$ (i.e., $\frac{\partial L}{\partial {\bf X}^l}$) and {\em (ii)} computation of loss gradient with respect to weight tensor ${\bf W}^l$ (i.e., $\frac{\partial L}{\partial {\bf W}^l}$). Standard implementations of these two components map the computation into Conv operation (i.e., the Conv operation during forward pass, shown in Fig.~\ref{fig:ConvLr}) by applying several tensor transformations. 

\noindent
\underline{\textit{Mapping $\frac{\partial L}{\partial {\bf X}^l}$ computation as Conv:}} Specifically, to compute $\frac{\partial L}{\partial {\bf X}^l}$, the following tensor transformations are applied:
\begin{itemize}
	\item The elements in each channel of the input loss gradient tensor (i.e., $\frac{\partial L}{\partial {\bf X}^{l+1}}$) are dilated\footnote{{\scriptsize Dilation is the process of inserting zeros in between the elements of a tensor, which expands its size.}} with ($S -$1) zeros, both horizontally and vertically.
	\item The gradient $\frac{\partial L}{\partial {\bf X}^{l+1}}$ is also padded with ($K_h-$1) zeros at the top/bottom and ($K_w-$1) zeros at the left/right.
	\item The elements in each channel of the weight tensor (i.e., ${\bf W}^l$) are flipped horizontally and vertically.
	\item ifmap and ofmap channel dimensions are interchanged.
\end{itemize}
After these transformations, the computation of $\frac{\partial L}{\partial {\bf X}^l}$ (acts as {\bf ofmap}) becomes a Conv operation with stride one (i.e., $S$ = 1) between the transformed $\frac{\partial L}{\partial {\bf X}^{l+1}}$ (acts as {\bf ifmap}) and ${\bf W}^l$ (acts as {\bf filter}) tensors.

\noindent
\underline{\textit{Mapping $\frac{\partial L}{\partial {\bf W}^l}$ computation as Conv:}} The mapping of $\frac{\partial L}{\partial {\bf W}^l}$ as Conv involves the following tensor transformations:
\begin{itemize}
	\item The transformation described at the first bullet for $\frac{\partial L}{\partial {\bf X}^l}$ computation is applied in $\frac{\partial L}{\partial {\bf X}^{l+1}}$ tensor.
	\item ifmap channel and batch dimensions are interchanged.
\end{itemize}
After these transformations, the computation of $\frac{\partial L}{\partial {\bf W}^l}$ (acts as {\bf ofmap}) becomes a Conv operation with stride one between the transformed $\frac{\partial L}{\partial {\bf X}^{l+1}}$ ({\bf filter}) and ${\bf X}^l$ ({\bf ifmap}) tensors.

\begin{figure}[t]
	\vspace{-0mm}
	\centering
	\includegraphics[width=3.5in]{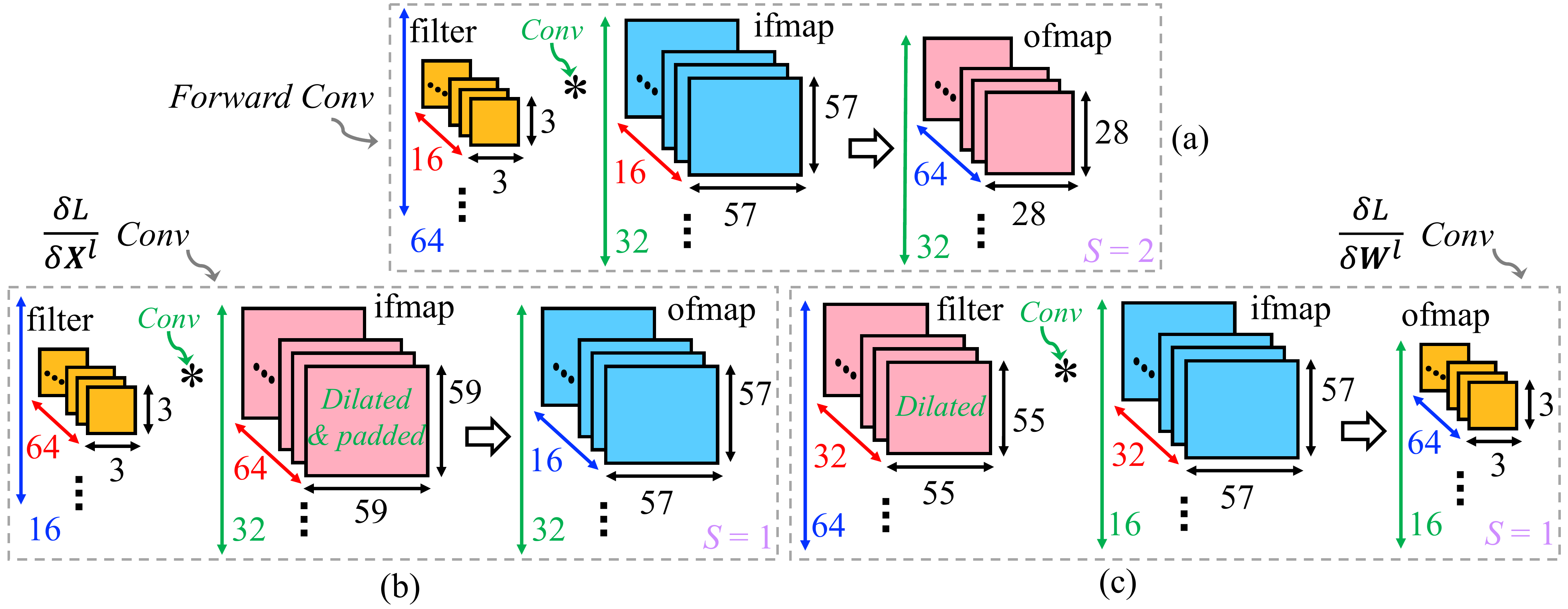}
	\caption{Numerical example of Conv dimensions during training: (a) Forward Conv for the $l^{th}$ layer, (b) Backward Conv to compute $\frac{\partial L}{\partial {\bf X}^l}$, and (c) Backward Conv to compute $\frac{\partial L}{\partial {\bf W}^l}$.}
	\label{fig:TrainConv_Ex}
	\vspace{-0mm}
\end{figure}

The simulation framework of SimDIT requires only the size of each tensor of a Conv layer to predict the associated hardware {\em performance statistics}. Therefore, from the size of each tensor of a Conv layer during the forward pass (i.e., forward Conv), we calculate the sizes of the transformed tensors and supply that as inputs to SimDIT to determine the data access and cycle count costs for executing $\frac{\partial L}{\partial {\bf X}^l}$ and $\frac{\partial L}{\partial {\bf W}^l}$. Since the computations are Conv operations, we use our models from Sections \ref{sec:DataModelConv} and \ref{sec:CycleModelConv}. The precise formulation to obtain the tensor sizes for the Conv operations corresponding to $\frac{\partial L}{\partial {\bf X}^l}$ and $\frac{\partial L}{\partial {\bf W}^l}$ (i.e., backward Conv operations) from the tensor sizes of the forward Conv is summarized in Table~\ref{tbl:ConvTrain} and Fig.~\ref{fig:TrainConv_Ex}. 

\ignore{
	\begin{table}[htb]
		\vspace{-0mm}
		\centering
		\caption{Conv Train}
		\label{tbl:ConvTrain}
		\vspace{-0mm}
		{\scriptsize
			\begin{tabular}{|c|l|l|}
				\hline
				& \multicolumn{1}{c|}{Filter dimensions, stride}      & \multicolumn{1}{c|}{Additional ifmap/ofmap dimensions}    \\ \hline
				\begin{tabular}[c]{@{}c@{}}For\\ $\frac{\partial L}{\partial {\bf X}^l}$\end{tabular} 
				& \begin{tabular}[c]{@{}l@{}} $K^B_h$ = $K^F_h$ \\ $K^B_w$ = $K^F_w$  \\ $IC^B$ = $OC^F$ \\ $OC^B$ = $IC^F$ \\ $S^B$ = 1\end{tabular} 
				& \begin{tabular}[c]{@{}l@{}}$OH^B$ = $IH^F$ \\ $OW^B$ = $IW^F$\\ $IH^B$ = $S^F(OH^F-1)+1+2(K^F_h-1)$ \\ $IW^B$ = $S^F(OW^F-1)+1+ 2(K^F_w-1)$ \\ 
					$N$ = $N$ \end{tabular} \\ \hline
				\begin{tabular}[c]{@{}c@{}}For\\ $\frac{\partial L}{\partial {\bf W}^l}$\end{tabular} 
				& \begin{tabular}[c]{@{}l@{}} $K^B_h$ = $S^F(OH^F-1)+1$\\ $K^B_w$ = $S^F(OW^F-1)+1$\\ $IC^B$ = $N^F$ \\ $OC^B$ = $OC^F$ \\ $S^B$ = 1\end{tabular} 
				& \begin{tabular}[c]{@{}l@{}} $OH^B$ = $K^F_h$\\ $OW^B$ = $K^F_w$\\ $IH^B$ = $IH^F$ \\ $IW^B$ = $IW^F$ \\ $N^B$ = $IC^F$\end{tabular}   \\ \hline
			\end{tabular}
		}
		\vspace{-0mm}
	\end{table}
}

\begin{table}[t]
	\vspace{-0mm}
	\centering
	\caption{Formulas showing the relationship among the dimension sizes of the forward Conv and backward Conv operations$^*$.}
	\label{tbl:ConvTrain}
	\vspace{-0mm}
	{\scriptsize
		\begin{tabular}{|cl|}
			\hline
			\multicolumn{1}{|c|}{Filter dimensions and stride}    & \multicolumn{1}{c|}{Additional ifmap/ofmap dimensions}    \\ \hline
			\multicolumn{2}{|c|}{For ${\partial L}/{\partial {\bf X}^l}$ computation as Conv}     \\ \hline
			\multicolumn{1}{|l|}{\begin{tabular}[c]{@{}l@{}} $K^B_h$ = $K^F_h$ \\ $K^B_w$ = $K^F_w$  \\ $IC^B$ = $OC^F$ \\ $OC^B$ = $IC^F$ \\ $S^B$ = 1 \end{tabular}}
			& \begin{tabular}[c]{@{}l@{}} $OH^B$ = $IH^F$ \\ $OW^B$ = $IW^F$\\ $IH^B$ = $S^F(OH^F-1)+1+2(K^F_h-1)$ \\ 
				$IW^B$ = $S^F(OW^F-1)+1+ 2(K^F_w-1)$ \\ $N$ = $N$ \end{tabular} \\ \hline
			\multicolumn{2}{|c|}{For ${\partial L}/{\partial {\bf W}^l}$ computation as Conv}   \\ \hline
			\multicolumn{1}{|l|}{\begin{tabular}[c]{@{}l@{}} $K^B_h$ = $S^F(OH^F-1)+1$\\ $K^B_w$ = $S^F(OW^F-1)+1$\\ $IC^B$ = $N^F$ \\ $OC^B$ = $OC^F$ \\ 
					$S^B$ = 1 \end{tabular}} 
			& \begin{tabular}[c]{@{}l@{}} $OH^B$ = $K^F_h$\\ $OW^B$ = $K^F_w$\\ $IH^B$ = $IH^F$ \\ $IW^B$ = $IW^F$ \\ $N^B$ = $IC^F$ \end{tabular}    \\ \hline
			\multicolumn{2}{l}{\textit{*superscript $F$ and $B$ indicate dimension sizes for the forward and}} \\
			\multicolumn{2}{l}{\textit{backward Conv, respectively.}} 
		\end{tabular}		
	}
\end{table}

\subsection{BN Loss Gradients: SIMD Array Operations (non-Conv)}
\label{sec:LossGradBN}

\begin{figure}[bt]
	\vspace{-0mm}
	\centering
	\includegraphics[width=3.0in]{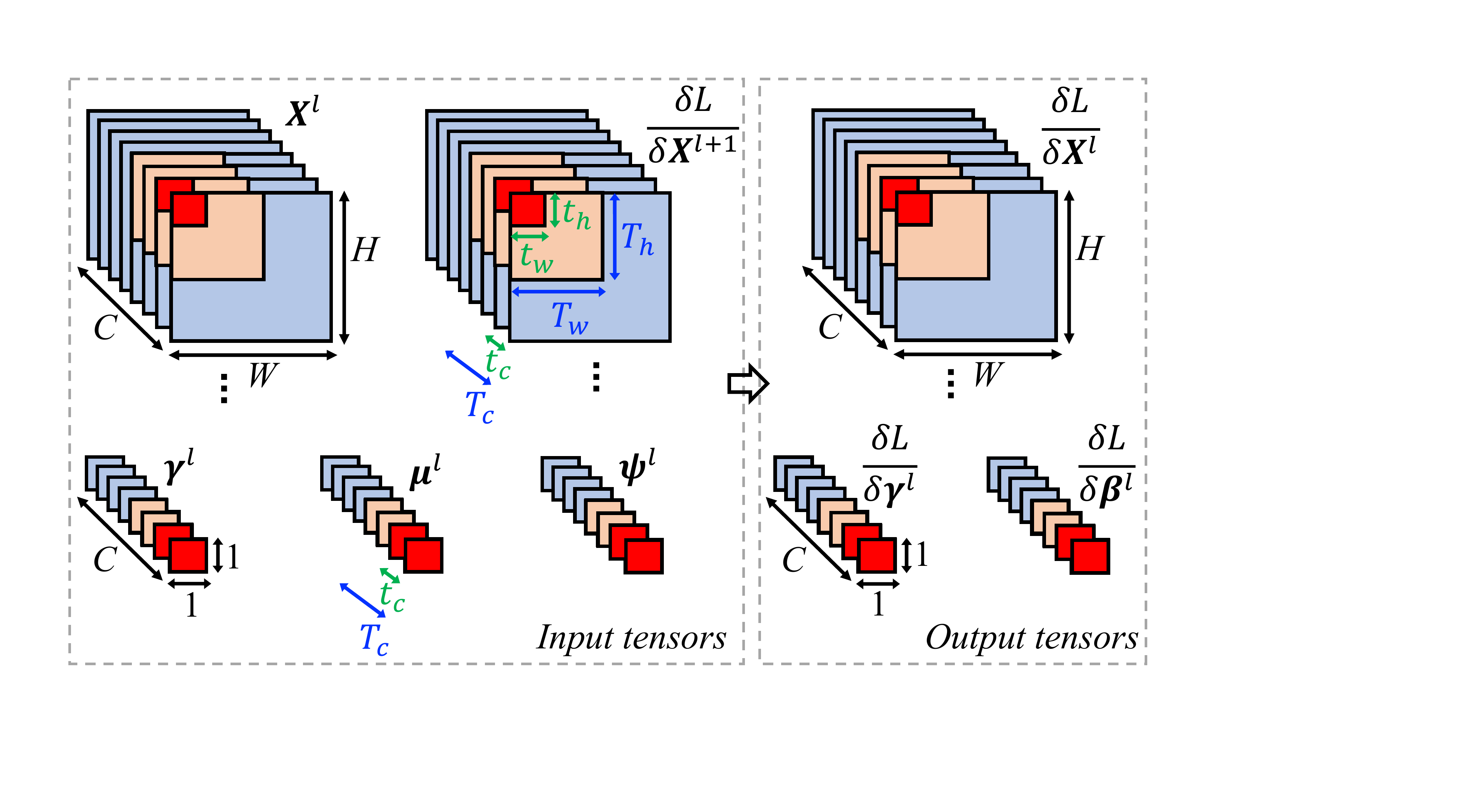}
	\vspace{-0mm}
	\caption{Illustration of input and output tensors including their notations during the backward pass through $l^{th}$ BN layer. The batch dimension of ${\bf X}^l$, $\frac{\partial L}{\partial {\bf X}^{l+1}}$, and $\frac{\partial L}{\partial {\bf X}^l}$ tensors and the associated notations (i.e., batch size: $N$; tiling parameters: $T_n$ and $t_n$) are omitted in the figure for better visualization.}
	\label{fig:BNGradTile}
\end{figure}

\noindent
As a representative non-Conv training operation, we present the modeling approach of SimDIT for batch normalization (BN) layer during backward pass. A similar modeling approach is adopted for all the non-Conv training operations listed in Table~\ref{tbl:LayerList}. In the forward pass of $l^{th}$ BN layer, computation using an input tensor, ${\bf X}^l$, of size ($H \times W \times N \times C$) and a pair of parameter tensors, scale ($\pmb {\gamma}^l$) and shift ($\pmb {\beta}^l$) each of size $C$, produces BN transformed output. The backward pass through $l^{th}$ BN layer (henceforth, we call it BN$_{back}$) consists of computing three components:
\begin{itemize}
	\item loss gradient with respect to parameter scale, $\frac{\partial L}{\partial \pmb {\gamma}^l}$ 
	\item loss gradient with respect to parameter shift, $\frac{\partial L}{\partial \pmb {\beta}^l}$
	\item loss gradient with respect to input, $\frac{\partial L}{\partial {\bf X}^l}$
\end{itemize}
\noindent
The left box of Fig.~\ref{fig:BNGradTile} shows the five tensors that are used as inputs for BN$_{back}$. The tensor $\pmb {\mu}^l$ denotes the batch mean and the tensor $\pmb {\psi}^l$ represents the inverse square root of batch variance. Both of these tensors are computed during the forward pass of BN layer and stored off-chip to be used during the backward pass. The tensor $\frac{\partial L}{\partial {\bf X}^{l+1}}$ indicates the input loss gradient. The right box shows the three output tensors of BN$_{back}$. The dimensions of all the input and output tensors are visually shown in the figure. The interpretation of the tiling template is same as Fig.~\ref{fig:TensorAddTile} (i.e., $T_{\phi}$ and $t_{\phi}$ stand for the sizes of outer and inner tile, respectively, for the dimension $\phi$ where $\phi \in \lbrace h, w, n, c\rbrace$). Outer tiles of all the input and output tensors are stored in VMem while the $K$ ALUs of the SIMD array operate in parallel to process inner tile computations every cycle using $t_h$ = $t_w$ = $t_n$ = 1, and $t_c$ = $K$.

The analytical expressions~\cite{BatchNorm2015} to compute the three outputs of BN$_{back}$ are shown in Equations~\eqref{eq:scale_grad}-\eqref{eq:data_grad}
where following standard implementations, analytical simplification is applied to obtain the expression for $\frac{\partial L}{\partial {\bf X}^l}$. The computations of the loss gradients use normalized form of the input tensor which is denoted by ${\bf \widehat{X}}^l$ and computed using~\eqref{eq:xhat_comp}. In BN$_{back}$, the operations inside nested $h, w, n$ loops are same where these three dimensions act as the batch dimension with an effective batch size of $N_{eff}$ = ($H \times W \times N$). Therefore, for notational clarity in Equations~\eqref{eq:xhat_comp}-\eqref{eq:data_grad}, we use two indices to indicate individual elements of each tensor (i.e., $i$ to index the elements along the $h, w, n$ loops while $j$ to index the elements along the $c$ loop). Besides, non-bold lower case notations are used to denote individual elements in each tensor, for example, $x^l_{ij}$ indicates an individual element in tensor ${\bf X}^l$.
%
\begin{align}
\hat{x}^l_{ij} =& \; (x^l_{ij} - \mu^l_j) \cdot \psi^l_j 
\label{eq:xhat_comp} \\
\frac{\partial L}{\partial \gamma^l_j} =& \; \sum_{i=1}^{N_{eff}} \Big( \frac{\partial L}{\partial x^{l+1}_{ij}} \cdot \hat{x}^l_{ij} \Big) 
\label{eq:scale_grad} \\
\frac{\partial L}{\partial \beta^l_j} =& \; \sum_{i=1}^{N_{eff}} \frac{\partial L}{\partial x^{l+1}_{ij}} 
\label{eq:shift_grad} \\
\frac{\partial L}{\partial x^l_{ij}} =& \; \frac{\gamma^l_j \cdot \psi^l_j}{N_{eff}} \Big(N_{eff} \cdot \frac{\partial L}{\partial x^{l+1}_{ij}} - \frac{\partial L}{\partial \gamma^l_j} \cdot \hat{x}^l_{ij} - \frac{\partial L}{\partial \beta^l_j}\Big)
\label{eq:data_grad}
\end{align}

\begin{algorithm}[t]
    {\scriptsize
	\textbf{INPUT:} Input tensors: ${\bf X}^l$, $\frac{\partial L}{\partial {\bf X}^{l+1}}$, $\pmb {\gamma}^l$, $\pmb {\mu}^l$, $\pmb {\psi}^l$; \\
	Layer shape parameters: $H$, $W$, $N$, $C$; \\
	Tiling parameters: $T_h$, $T_w$, $T_n$, $T_c$. \\	
	\textbf{OUTPUT:} Loss gradient tensors: $\frac{\partial L}{\partial {\bf X}^l}$,  $\frac{\partial L}{\partial \pmb {\gamma}^l}$, $\frac{\partial L}{\partial \pmb {\beta}^l}$.
	\begin{algorithmic} [1]
		\For{ each tile along $c$ loop}  \label{algo1:cloop_start}
			\Statex \hspace{0.35cm} {\em // Computing loss gradients with respect to scale and shift:}
			\State Load tiles of $\pmb {\mu}^l$ and $\pmb {\psi}^l$   \label{algo1:load_mu_istd}
			\For{ each tile along $n$ loop}  \label{algo1:nloop_start1}
				\For{ each tile along $w$ loop}  \label{algo1:hloop_start1}
					\For{ each tile along $h$ loop}  \label{algo1:wloop_start1}
					
						\State Load tiles of ${\bf X}^l$ and $\frac{\partial L}{\partial {\bf X}^{l+1}}$  \label{algo1:load_data_grad}
						\State Compute tile of ${\bf \widehat{X}}^l$ ({\em require sub and mul}) \label{algo1:comp_xhat} 
						\State Compute partial sums for tiles of $\frac{\partial L}{\partial \pmb {\gamma}^l}$ and $\frac{\partial L}{\partial \pmb {\beta}^l}$ \label{algo1:comp_param_grad} 
						\Statex \hspace{1.9cm}({\em require mul and add})
						\State Store tile of ${\bf \widehat{X}}^l$ \label{algo1:store_xhat} 
					
					\EndFor  \label{algo1:wloop_end1}
				\EndFor  \label{algo1:hloop_end1}
			\EndFor  \label{algo1:nloop_end1}
			
			\Statex \hspace{0.35cm} {\em // Computing loss gradient with respect to input:}
			\State Load tile of $\pmb {\gamma}^l$ \label{algo1:load_scale}
			\State Compute the term outside the parenthesis in~\eqref{eq:data_grad}  \label{algo1:comp_cons}
			\Statex \hspace{0.38cm} ({\em require mul and div}) 
			\For{ each tile along $n$ loop}  \label{algo1:nloop_start2}
				\For{ each tile along $w$ loop}  \label{algo1:hloop_start2}
					\For{ each tile along $h$ loop}  \label{algo1:wloop_start2}
						
						\State Load tiles of ${\bf \widehat{X}}^l$ and $\frac{\partial L}{\partial {\bf X}^{l+1}}$  \label{algo1:load_xhat_grad}
						\State Compute tile of $\frac{\partial L}{\partial {\bf X}^l}$ ({\em require mul and sub}) \label{algo1:comp_out_grad}
						\State Store tile of $\frac{\partial L}{\partial {\bf X}^l}$  \label{algo1:store_out_grad}
						
					\EndFor  \label{algo1:wloop_end2}
				\EndFor  \label{algo1:hloop_end2}
			\EndFor  \label{algo1:nloop_end2}
			
			\State Store tiles of $\frac{\partial L}{\partial \pmb {\gamma}^l}$ and $\frac{\partial L}{\partial \pmb {\beta}^l}$  \label{algo1:store_param_grad} 
					
		\EndFor  \label{algo1:cloop_end}
				
	\end{algorithmic}
	\caption{Steps to Compute Loss Gradients for $l^{th}$ BN Layer}
	\label{alg:BN_grad_algo}
    }
\end{algorithm}

Unlike most non-Conv layers where the computation comprises one or two kinds of arithmetic operations, the computation of BN$_{back}$ involves several types of arithmetic operations that are embedded within multiple nested loops. Consequently, there are many ways to map the computation of BN$_{back}$ in the generic SIMD processor. In SimDIT, we efficiently schedule the BN$_{back}$ computations in the SIMD array to minimize the number of cycle counts as well as data access cost for the BN$_{back}$ layer. 
Algorithm~\ref{alg:BN_grad_algo} summarizes the steps adopted by SimDIT to compute the three loss gradients of the BN$_{back}$ layer. 
The algorithm shows the computation steps using outer tiles only, which are referred to simply as tiles. 

For each channel, the loss gradient computation with respect to input requires the computed loss gradients with respect to scale and shift (Equation~\eqref{eq:data_grad}). Hence, as shown in Algorithm~\ref{alg:BN_grad_algo}, we finish computing tiles of $\frac{\partial L}{\partial \pmb {\gamma}^l}$ and $\frac{\partial L}{\partial \pmb {\beta}^l}$ before $\frac{\partial L}{\partial {\bf X}^l}$. The BN$_{back}$ computations proceed in two parts:
\begin{enumerate}[label=(\roman*)]
	\item Part-1 (Lines~\ref{algo1:cloop_start}--\ref{algo1:nloop_end1}, \ref{algo1:store_param_grad}, \ref{algo1:cloop_end}) that computes the $\frac{\partial L}{\partial \pmb {\gamma}^l}$ and $\frac{\partial L}{\partial \pmb {\beta}^l}$ tensors using Equations~\eqref{eq:xhat_comp}--\eqref{eq:shift_grad}.
	\item Part-2 (Lines~\ref{algo1:cloop_start}, \ref{algo1:load_scale}--\ref{algo1:nloop_end2}, \ref{algo1:cloop_end}) that computes the $\frac{\partial L}{\partial {\bf X}^l}$ tensor using~\eqref{eq:data_grad}.
\end{enumerate}

\noindent The processing of Part-1 involves loading tiles of respective input tensors from DRAM (Lines~\ref{algo1:load_mu_istd} and \ref{algo1:load_data_grad}). After computing a tile of ${\bf \widehat{X}}^l$, it is stored back to DRAM to be used by Part-2 (Line~\ref{algo1:store_xhat}). Since ${\bf \widehat{X}}^l$ is a 4D tensor, without this writeback to DRAM, the data will be lost as the computation proceeds along the $h$, $w$, and $n$ loops. For the 1D tensors of $\frac{\partial L}{\partial \pmb {\gamma}^l}$ and $\frac{\partial L}{\partial \pmb {\beta}^l}$, the accumulation across the $h$, $w$, and $n$ loops gets completed in Line~\ref{algo1:nloop_end1} and the data for the corresponding loss gradient tiles are formed. Under the same $c$ loop, these completed tiles of $\frac{\partial L}{\partial \pmb {\gamma}^l}$ and $\frac{\partial L}{\partial \pmb {\beta}^l}$ are reused in Part-2 to compute $\frac{\partial L}{\partial {\bf X}^l}$ before they are stored to DRAM (Line~\ref{algo1:store_param_grad}), reducing off-chip accesses.

During the processing of Part-2, tiles of $\pmb {\gamma}^l$, ${\bf \widehat{X}}^l$, and $\frac{\partial L}{\partial {\bf X}^{l+1}}$ are loaded from DRAM (Lines~\ref{algo1:load_scale} and \ref{algo1:load_xhat_grad}). After computing a tile of the 4D tensor $\frac{\partial L}{\partial {\bf X}^l}$, it is stored back to DRAM (Line~\ref{algo1:store_out_grad}). This is repeated until all three output tensors are computed entirely. The computation of Part-1 and Part-2 requires a series of arithmetic addition (add), subtraction (sub), multiplication (mul), and division (div) operations (Lines~\ref{algo1:comp_xhat}, \ref{algo1:comp_param_grad}, \ref{algo1:comp_cons}, and \ref{algo1:comp_out_grad}).

Using the scheduling steps in Algorithm~\ref{alg:BN_grad_algo}, we develop models for DRAM accesses, SRAM accesses, computation cycles and DRAM stall cycles for the BN$_{back}$ layer. The derivations for these models are presented in Appendix~\ref{sec:BNBack_Models}.

\section{Energy and Power Computation}
\label{sec:EnPow}

\noindent
We combine the performance statistics (i.e., cycle counts and number of accesses) from SimDIT with post SP\&R {power-performance characteristics} for a hardware configuration to compute end-to-end execution energy and power of a network. We use the following hardware data from SP\&R:
\begin{itemize}
	\item dynamic and leakage power of systolic PE array (SA) and SIMD array denoted as ${\cal P}_{SA-dyn}$, ${\cal P}_{SA-leak}$, ${\cal P}_{SIMD-dyn}$, and ${\cal P}_{SIMD-leak}$, respectively.
	\item per bit access energy for each on-chip buffer ($e_{buff}$) according to their size specifications.
	\item effective clock period ($T_{clk}$).
\end{itemize}

\noindent
The total energy to execute a network, $E_{total}$, is:
\begin{equation}
E_{total} = E_{SA} + E_{SIMD} + E_S + E_D
\label{eq:Etotal}
\end{equation}
where $E_{SA}$ and $E_{SIMD}$ indicate the energy consumption of the SA and SIMD array, respectively, as they compute a network while $E_S$ and $E_D$ represent data access energy from the on-chip SRAMs and off-chip DRAM, respectively. The energy of the SA and SIMD hardware components include dynamic energy as well as leakage energy and computed as:
\begin{align}
E_{SA} =& \; \left({\cal C}_{SA} \cdot {\cal P}_{SA-dyn} + {\cal L}_{total} \cdot {\cal P}_{SA-leak}\right) \cdot T_{clk}
\nonumber \\
E_{SIMD} =& \; \left({\cal C}_{SIMD} \cdot {\cal P}_{SIMD-dyn} + \right . \nonumber \\
& \hspace{2cm} \left . {\cal L}_{total} \cdot {\cal P}_{SIMD-leak}\right) \cdot T_{clk}
\label{eq:Esimd}
\end{align}

\noindent
Here, ${\cal C}_{SA}$ [${\cal C}_{SIMD}$] captures the number of cycles when the SA [SIMD] core actively performs computation while ${\cal L}_{total}$ denotes end-to-end latency (i.e., total number of cycles to execute a network that includes computation as well as stall cycles of the SA and SIMD components). The data access energy from the on-chip SRAMs and off-chip DRAM is:
\begin{equation}
E_S = \textstyle \sum_{buff} \left(A_{S_{buff}} \cdot e_{buff}\right) \; \; ; \; \; 
E_D = A_{D_{total}} \cdot e_D
\label{eq:Esram_dram}
\end{equation}
Here, $A_{S_{buff}}$ is the total number of accesses for a buffer $buff \in \{\mbox{WBuf, IBuf, OBuf, BBuf, VMem, IMem}\}$, and $A_{D_{total}}$ is the total number of accesses from the DRAM with per bit access energy of $e_D$. Finally, the average power to execute a network is:
\begin{equation}
{\cal P}_{avg} = E_{total} / \left({\cal L}_{total} \cdot T_{clk}\right)
\label{eq:Power_avg}
\end{equation}
\section{Evaluation}
\label{sec:ResltSim}

\noindent
We use SimDIT to analyze the impact of Conv and non-Conv workloads on end-to-end performance of a CNN. We also perform design space exploration to obtain optimized accelerator design solutions. As representative networks, we use ResNet-50, ResNet-18, VGG16, and AlexNet. These are widely used CNN topologies that consist of a diverse set of layers (i.e., Conv, FC, ReLU, Tensor-add, max Pool, global average Pool, and BN). 
For inference workload, we obtain the network graph containing the tensor specifications of each layer using models from PyTorch (ResNet-50, ResNet-18) and ONNX Model Zoo (VGG16, AlexNet). For training workload, we traverse the inference graph to obtain the tensor specifications of each layer for the forward and backward pass of training. We develop a tiling generator that generates valid tiling parameters for each type of layer using the configuration of the hardware (i.e., the {\em Hardware Specifications}, summarized in Table~\ref{tbl:HrdParam}). Since the interface of SimDIT uses tiling parameters as input, they can also be provided externally by the user (for example, valid tiling parameters generated by an external compiler).

To obtain the backend hardware data, we use RTL of GeneSys from VeriGOOD-ML open-source repository~\cite{VerigoodML_github}, an automated Verilog generator for ASIC-based DNN accelerator. We perform backend synthesis place-and-route of various GeneSys designs (both inference and training hardware designs) using a commercial 12nm technology node to obtain the post SP\&R power performance characteristics of the hardware. We use a commercial memory compiler to obtain the per bit access energy of the on-chip SRAMs. The per bit access energy of HBM2 DRAM memory is obtained from~\cite{OConnor2017}.

\begin{table}[htb]
	\vspace{-0mm}
	\centering
	\caption{Percentage of runtime, energy, and number of data accesses for the non-Conv layers with respect to total network performance (i.e., Conv + non-Conv) for ResNet-50 during training and inference.}
	\label{tbl:conv_non_conv}
	\vspace{-0mm}
	{\scriptsize
		\begin{tabular}{|ccccccc|}
			\hline
			\multicolumn{1}{|c|}{$J$$\times$$K$}   & \multicolumn{1}{c|}{\begin{tabular}[c]{@{}c@{}}non-Conv\\ runtime\end{tabular}} & \multicolumn{1}{c|}{\begin{tabular}[c]{@{}c@{}}non-Conv\\ on-chip\\ access\end{tabular}} & \multicolumn{1}{c|}{\begin{tabular}[c]{@{}c@{}}non-Conv\\ off-chip\\ access\end{tabular}} & \multicolumn{1}{c|}{\begin{tabular}[c]{@{}c@{}}non-Conv\\ energy\end{tabular}} & \multicolumn{1}{c|}{\begin{tabular}[c]{@{}c@{}}Total\\ power\end{tabular}} & \begin{tabular}[c]{@{}c@{}}Total\\ runtime\end{tabular} \\ \hline
			\multicolumn{7}{|c|}{Training (per iteration)}                                                                                                                                                                                                                                                                                                                                                                                                                                                               \\ \hline
			\multicolumn{1}{|c|}{16$\times$16} & \multicolumn{1}{c|}{41.9\%}  & \multicolumn{1}{c|}{4.1\%}     & \multicolumn{1}{c|}{44.8\%}     & \multicolumn{1}{c|}{50.3\%}     & \multicolumn{1}{c|}{0.63W}   & 6.86s       \\ \hline
			\multicolumn{1}{|c|}{32$\times$32} & \multicolumn{1}{c|}{56.6\%}      & \multicolumn{1}{c|}{4.1\%}    & \multicolumn{1}{c|}{59.3\%}     & \multicolumn{1}{c|}{52.3\%}   & \multicolumn{1}{c|}{1.58W}     & 2.54s        \\ \hline
			\multicolumn{1}{|c|}{64$\times$64} & \multicolumn{1}{c|}{59.5\%}    & \multicolumn{1}{c|}{2.7\%}     & \multicolumn{1}{c|}{56.2\%}    & \multicolumn{1}{c|}{49.4\%}              & \multicolumn{1}{c|}{3.94W}               & 1.21s    \\ \hline
			\multicolumn{7}{|c|}{Inference (per image)}                                                                                                                                                                                                                                                                                                                                                                                                                                                                  \\ \hline
			\multicolumn{1}{|c|}{16$\times$16} & \multicolumn{1}{c|}{30.1\%}    & \multicolumn{1}{c|}{1.9\%}  & \multicolumn{1}{c|}{38.7\%}     & \multicolumn{1}{c|}{33.2\%}             & \multicolumn{1}{c|}{0.46W}             & 0.028s         \\ \hline
			\multicolumn{1}{|c|}{32$\times$32} & \multicolumn{1}{c|}{41.6\%}    & \multicolumn{1}{c|}{2.0\%}  & \multicolumn{1}{c|}{54.4\%}   & \multicolumn{1}{c|}{40.3\%}             & \multicolumn{1}{c|}{1.02W}              & 0.010s           \\ \hline
			\multicolumn{1}{|c|}{64$\times$64} & \multicolumn{1}{c|}{49.3\%}   & \multicolumn{1}{c|}{1.8\%}   & \multicolumn{1}{c|}{56.6\%}    & \multicolumn{1}{c|}{38.2\%}             & \multicolumn{1}{c|}{2.55W}               & 0.004s         \\ \hline
		\end{tabular}		
	}
	\vspace{-0mm}
\end{table}

\subsection{Impact of Conv and non-Conv Operations}
\label{sec:ConvNonConv}

\noindent
Table~\ref{tbl:conv_non_conv} demonstrates the fraction of runtime, energy, and number of on-chip and off-chip accesses for the non-Conv layers of ResNet-50 with respect to the total network performance across different hardware configurations. The table also presents the total power and runtime to execute the network (last two columns). 

For training, the evaluations are shown for a single iteration through the network (i.e., a forward pass + a backward pass) using a batch size of 32 on three hardware configurations: \\
{\bf HT1} Array size $=$ 16$\times$16 \\
\hspace*{2mm} $[Wbuf, Ibuf, Obuf, Vmem]$ $=$ [256, 128, 256, 256] kB \\
\hspace*{2mm} Bandwidth per off-chip interface $=$ 128 bits/cycle \\
{\bf HT2} Array size $=$ 32$\times$32 \\
\hspace*{2mm} $[Wbuf, Ibuf, Obuf, Vmem]$ $=$ [512, 256, 512, 512] kB\\
\hspace*{2mm} Bandwidth per off-chip interface $=$ 256 bits/cycle \\
{\bf HT3} Array size $=$ 64$\times$64 \\
\hspace*{2mm} $[Wbuf, Ibuf, Obuf, Vmem]$ $=$ [1024, 512, 1024, 1024] kB \\
\hspace*{2mm} Bandwidth per off-chip interface $=$ 512 bits/cycle \\
Following~\cite{TPUv2-2020}, across all three hardware substrates, we use 16 bits for ifmap/weight and 32 bits for psum data on the systolic PE array, and 32 bits for all input/output data of the SIMD array. From Table~\ref{tbl:conv_non_conv}, across all three hardware configurations, the energy and runtime spent to execute the non-Conv layers are a significant fraction of the network performance e.g., up to 59.5\% runtime and 52.3\% energy are consumed by the non-Conv layers. This is not surprising since the training workload comprises a large set of non-Conv layers. As listed in Table~\ref{tbl:LayerList}, there are four types of non-Conv layers during the forward pass of training while it consists of five categories of non-Conv layers during the backward pass.

For the inference workload, the evaluations in Table~\ref{tbl:conv_non_conv} are shown for a single inference using a batch size of 1. Similar to~\cite{Jouppi2017}, the inference hardware uses 8-bit for ifmap/weight and 32-bit for psum on the SA component while 32-bit data on the SIMD component. The inference results are shown using three hardware configurations with different array sizes, SRAM specifications, and off-chip bandwidths: \\
{\bf HI1} Array size $=$ 16$\times$16 \\
\hspace*{2mm} $[Wbuf, Ibuf, Obuf, Vmem]$ $=$ [32, 32, 128, 128] kB \\
\hspace*{2mm} Bandwidth per off-chip interface $=$ 128 bits/cycle \\
{\bf HI2} Array size $=$ 32$\times$32 \\
\hspace*{2mm} $[Wbuf, Ibuf, Obuf, Vmem]$ $=$ [256, 128, 512, 512] kB\\
\hspace*{2mm} Bandwidth per off-chip interface $=$ 256 bits/cycle \\
{\bf HI3} Array size $=$ 64$\times$64 \\
\hspace*{2mm} $[Wbuf, Ibuf, Obuf, Vmem]$ $=$ [512, 256, 1024, 1024] kB \\
\hspace*{2mm} Bandwidth per off-chip interface $=$ 512 bits/cycle \\
It is evident that the runtime and energy overheads from the non-Conv layers are comparable to that of Conv layers. While the overheads are smaller for HI1, with larger array size, as the ratio of the computing elements (i.e., \#of PEs vs. \#of ALUs) in the 2D SA vs. 1D SIMD component increases, the performance overheads from non-Conv layers increases, e.g., 49.3\% runtime and 38.2\% energy of HI3 for non-Conv layers. 

For a similar analysis on ResNet-18, Table~\ref{tbl:conv_non_conv_r18} summarizes the fraction of runtime, energy, on-chip, and off-chip accesses for the non-Conv layers with respect to the total network performance while executed on HT1, HT2, HT3 for training and HI1, HI2, HI3 for inference. Similar trends are seen for ResNet-18 where non-Conv layers constitute comparable runtime and energy overheads to that of Conv layers: e.g., 17.4\%$-$45.4\% overhead in runtime while 22.0\%$-$42.7\% overhead in energy across the inference and training workloads.

\begin{table}[t]
	\vspace{-0mm}
	\centering
	\caption{Percentage of runtime, energy, and number of data accesses for the non-Conv layers with respect to total network performance for ResNet-18 during training and inference.}
	\label{tbl:conv_non_conv_r18}
	\vspace{-0mm}
	{\scriptsize
		\resizebox{1.0\linewidth}{!}{
		\begin{tabular}{|ccccccc|}
			\hline
			\multicolumn{1}{|c|}{$J$$\times$$K$}   & \multicolumn{1}{c|}{\begin{tabular}[c]{@{}c@{}}non-Conv\\ runtime\end{tabular}} & \multicolumn{1}{c|}{\begin{tabular}[c]{@{}c@{}}non-Conv\\ on-chip\\ access\end{tabular}} & \multicolumn{1}{c|}{\begin{tabular}[c]{@{}c@{}}non-Conv\\ off-chip\\ access\end{tabular}} & \multicolumn{1}{c|}{\begin{tabular}[c]{@{}c@{}}non-Conv\\ energy\end{tabular}} & \multicolumn{1}{c|}{\begin{tabular}[c]{@{}c@{}}Total\\ power\end{tabular}} & \begin{tabular}[c]{@{}c@{}}Total\\ runtime\end{tabular} \\ \hline
			\multicolumn{7}{|c|}{Training (per iteration)}                                                                                                                                                                                                                                                                                                                                                                                                                                                               \\ \hline
			\multicolumn{1}{|c|}{16$\times$16} & \multicolumn{1}{c|}{30.5\%}  & \multicolumn{1}{c|}{2.7\%}     & \multicolumn{1}{c|}{41.8\%}     & \multicolumn{1}{c|}{42.7\%}     & \multicolumn{1}{c|}{0.57W}   & 2.50s       \\ \hline
			\multicolumn{1}{|c|}{32$\times$32} & \multicolumn{1}{c|}{41.8\%}      & \multicolumn{1}{c|}{2.4\%}    & \multicolumn{1}{c|}{60.0\%}     & \multicolumn{1}{c|}{40.9\%}   & \multicolumn{1}{c|}{1.56W}     & 0.91s        \\ \hline
			\multicolumn{1}{|c|}{64$\times$64} & \multicolumn{1}{c|}{45.4\%}    & \multicolumn{1}{c|}{1.5\%}     & \multicolumn{1}{c|}{56.1\%}    & \multicolumn{1}{c|}{36.9\%}              & \multicolumn{1}{c|}{4.22W}               & 0.42s    \\ \hline
			\multicolumn{7}{|c|}{Inference (per image)}                                                                                                                                                                                                                                                                                                                                                                                                                                                                  \\ \hline
			\multicolumn{1}{|c|}{16$\times$16} & \multicolumn{1}{c|}{17.4\%}    & \multicolumn{1}{c|}{1.2\%}  & \multicolumn{1}{c|}{31.3\%}     & \multicolumn{1}{c|}{24.1\%}             & \multicolumn{1}{c|}{0.36W}             & 0.011s         \\ \hline
			\multicolumn{1}{|c|}{32$\times$32} & \multicolumn{1}{c|}{24.7\%}    & \multicolumn{1}{c|}{1.1\%}  & \multicolumn{1}{c|}{46.0\%}   & \multicolumn{1}{c|}{26.6\%}             & \multicolumn{1}{c|}{0.91W}              & 0.004s           \\ \hline
			\multicolumn{1}{|c|}{64$\times$64} & \multicolumn{1}{c|}{30.0\%}   & \multicolumn{1}{c|}{0.9\%}   & \multicolumn{1}{c|}{46.5\%}    & \multicolumn{1}{c|}{22.0\%}             & \multicolumn{1}{c|}{2.64W}               & 0.002s         \\ \hline
		\end{tabular}		
		}
	}
\end{table}

\begin{table}[b]
	\centering
	\caption{Performance comparison of optimal hardware allocation with worst-case allocation for ResNet-50 inference workload across four different array sizes.}
	\label{tbl:OptDist}
	\vspace{-0mm}
	{\scriptsize
		\begin{tabular}{|c|cc|c|}
			\hline
			\multirow{2}{*}{$J$$\times$$K$} & \multicolumn{2}{c|}{Optimal resource allocation}           & \multirow{2}{*}{\begin{tabular}[c]{@{}c@{}}Improvement\\ over \\ worst-case\\ allocation\end{tabular}} \\ \cline{2-3}
			& \multicolumn{1}{c|}{\begin{tabular}[c]{@{}c@{}}SARM sizes (kB)\\ {[}$Wbuf$, $Ibuf$, \\ $Obuf$, $Vmem${]}\end{tabular}} & \begin{tabular}[c]{@{}c@{}}Bandwidth (bits/cycle)\\ {[}$BW_w$, $BW_i$, \\ $BW_o$, $BW_v${]}\end{tabular} &     \\ \hline
			16$\times$16    & \multicolumn{1}{c|}{{[}128, 256, 64, 64{]}}    & {[}64, 64, 128, 256{]}   & 9.64$\times$                  \\ \hline
			32$\times$32    & \multicolumn{1}{c|}{{[}256, 256, 128, 256{]}}     & {[}128, 128, 256, 512{]}     & 14.45$\times$         \\ \hline
			64$\times$64       & \multicolumn{1}{c|}{{[}256, 512, 256, 1024{]}}       & {[}256, 256, 512, 1024{]}     & 18.43$\times$           \\ \hline
			128$\times$128      & \multicolumn{1}{c|}{{[}512, 512, 512, 2048{]}}       & {[}512, 512, 1024, 2048{]}    & 25.55$\times$          \\ \hline
		\end{tabular}
	}
	\vspace{-0mm}
\end{table}

\begin{table}[b]
	\vspace{-0mm}
	\centering
	\caption{Performance comparison of optimal hardware allocation with worst-case allocation for inference workload on a 64$\times$64 array using budget constraint pair of (2048kB, 2048 bits/cycle) across different networks.}
	\label{tbl:OptDist_addi}
	\vspace{-0mm}
	{\scriptsize
		\begin{tabular}{|c|cc|c|}
			\hline
			\multirow{2}{*}{Network} & \multicolumn{2}{c|}{Optimal resource allocation}           & \multirow{2}{*}{\begin{tabular}[c]{@{}c@{}}Improvement\\ over \\ worst-case\\ allocation\end{tabular}} \\ \cline{2-3}
			& \multicolumn{1}{c|}{\begin{tabular}[c]{@{}c@{}}SARM sizes (kB)\\ {[}$Wbuf$, $Ibuf$, \\ $Obuf$, $Vmem${]}\end{tabular}} & \begin{tabular}[c]{@{}c@{}}Bandwidth (bits/cycle)\\ {[}$BW_w$, $BW_i$, \\ $BW_o$, $BW_v${]}\end{tabular} &     \\ \hline
			ResNet-18    & \multicolumn{1}{c|}{{[}512, 128, 128, 1024{]}}    & {[}512, 256, 256, 1024{]}   & 13.85$\times$                  \\ \hline
			VGG16    & \multicolumn{1}{c|}{{[}512, 256, 512, 512{]}}     & {[}512, 256, 256, 1024{]}     & 19.94$\times$         \\ \hline
			AlexNet       & \multicolumn{1}{c|}{{[}1024, 256, 128, 512{]}}       & {[}512, 256, 256, 1024{]}     & 33.72$\times$           \\ \hline
		\end{tabular}
	}
	\vspace{-0mm}
\end{table}

\subsection{Design Space Exploration}
\label{sec:SimSDE}

\noindent
{\bf \textit{Optimal allocation of hardware resources:}} We use SimDIT to perform design space exploration (DSE) to optimally distribute various hardware resources for inference workload. Under specified budgets for total on-chip SRAM size and total available off-chip bandwidth, we exhaustively search the design space by varying 8 parameters (i.e., sizes and bandwidths for WBuf, IBuf, OBuf, and VMem) while all hardware points lie within 15\% deviation from the budget constraints to allow greater flexibility to find the optimum. We use 8-bit ifmap/weight and 32-bit psum for SA, and 32-bit data for the SIMD component.

Table~\ref{tbl:OptDist} shows the optimal distribution of hardware resources across four different sizes of $J$$\times$$K$ for ResNet-50 inference using budget constraint pair of (512kB, 512 bits/cycle), (1024kB, 1024 bits/cycle), (2048kB, 2048 bits/cycle), and (4096kB, 4096 bits/cycle) for the 16$\times$16, 32$\times$32, 64$\times$64, and 128$\times$128 array, respectively. The DSE uses number of total network execution cycles as the performance metric. The rightmost column shows the improvement of optimal hardware point as compared to the worst-case allocation, and thus, represents the best-case to worst-case performance range. The large performance range underscores the importance of optimal distribution of hardware resources, for example, if the hardware resource allocation is not optimized by careful analysis, there can be a performance penalty of up to 18.43$\times$  on the 64$\times$64 array. Similar performance penalty is observed for ResNet-18, VGG16, and AlexNet. For these three networks, optimal hardware parameters are shown in Table~\ref{tbl:OptDist_addi} for a 64$\times$64 array, where, for example, VGG16 demonstrates a performance penalty of up to 19.94$\times$.

\begin{figure}[t]
	\vspace{-0mm}
	\centering
	\includegraphics[width=3.0in]{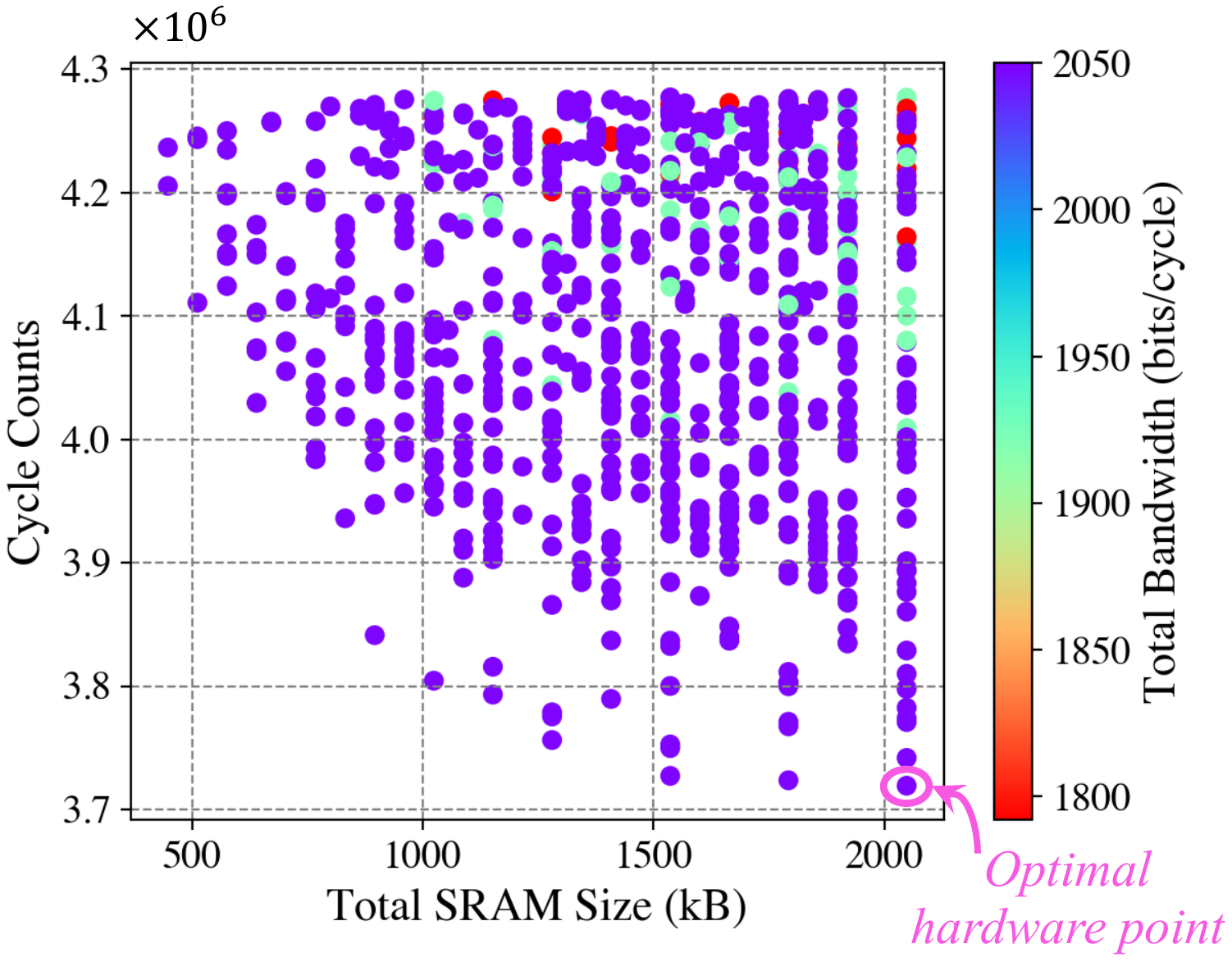}
	\vspace{-0mm}
	\caption{Distribution of cycle counts vs. total SRAM size and total bandwidth for various hardware points with maximum 15\% performance penalty with respect to optimal (64$\times$64 array, ResNet-50 inference workload). The third axis of the plot (i.e., total bandwidth) is indicated by color heatmap.}
	\label{fig:LandPlot}
	\vspace{-0mm}
\end{figure}

\begin{table}[t]
	\vspace{-0mm}
	\centering
	\caption{Performance comparison of optimal hardware point with economic design solutions for ResNet-50 inference on 64$\times$64 array.}
	\label{tbl:LandTable}
	\vspace{-0mm}
	{\scriptsize
		\begin{tabular}{|l|c|c|c|}
			\hline
			\begin{tabular}[c]{@{}l@{}}Hardware \\ point\end{tabular}     & \begin{tabular}[c]{@{}c@{}}Total SRAM size (kB)\\ {[}$Wbuf$, $Ibuf$, \\ $Obuf$, $Vmem${]}\end{tabular} & \begin{tabular}[c]{@{}c@{}}Total bandwidth (bits/cycle)\\ {[}$BW_w$, $BW_i$, \\ $BW_o$, $BW_v${]}\end{tabular} & \begin{tabular}[c]{@{}c@{}}Penalty \\ w.r.t. \\ optimal\end{tabular} \\ \hline
			Optimal         & 
			\begin{tabular}[c]{@{}c@{}}2048\\ {[}256, 512, 256, 1024{]}\end{tabular}     & \begin{tabular}[c]{@{}c@{}}2048\\ {[}256, 256, 512, 1024{]}\end{tabular}         & --          \\ \hline
			\begin{tabular}[c]{@{}l@{}}Minimized\\ SRAM size\end{tabular} & \begin{tabular}[c]{@{}c@{}}448\\ {[}128, 128, 128, 64{]}\end{tabular}      & \begin{tabular}[c]{@{}c@{}}2048\\ {[}256, 256, 512, 1024{]}\end{tabular}                           & 13.1\%                                                             \\ \hline
			\begin{tabular}[c]{@{}l@{}}Minimized\\ bandwidth\end{tabular} & \begin{tabular}[c]{@{}c@{}}1024\\ {[}256, 256, 256, 256{]}\end{tabular}                        & \begin{tabular}[c]{@{}c@{}}1792\\ {[}256, 256, 256, 1024{]}\end{tabular}                           & 14.6\%                                                             \\ \hline
		\end{tabular}
	}
\end{table}

\noindent
{\bf \textit{Design landscape analysis:}} We now analyze the flatness of the design space around the optimal hardware point for the 64$\times$64 array under the budget boundaries of 2048kB SRAM and 2048 bits/cycle off-chip bandwidth using ResNet-50. This analysis enables finding economic design solutions at the cost of small penalty in performance. Fig.~\ref{fig:LandPlot} shows the distribution of cycle counts for all the hardware points for which the performances are within 15\% of the optimal cycle counts. It is clear that there are many design points that deliver performance close to the optimal using a smaller hardware resource (i.e., on-chip memory size and/or off-chip bandwidth) as compared to the optimal hardware point.

Table~\ref{tbl:LandTable} presents the performance comparison of the optimal hardware point with two economic design solutions that are obtained from the hardware points in Fig.~\ref{fig:LandPlot}. The design that minimizes the on-chip SRAM size saves 78.1\% memory cost over the optimum with only a 13.1\% penalty in performance. Another design point that minimizes the off-chip bandwidth uses 50\% less memory and 12.5\% less bandwidth as compared to the optimal point with only a 14.6\% penalty in performance.


\begin{figure}[t]
	\centering
	\includegraphics[width=2.8in]{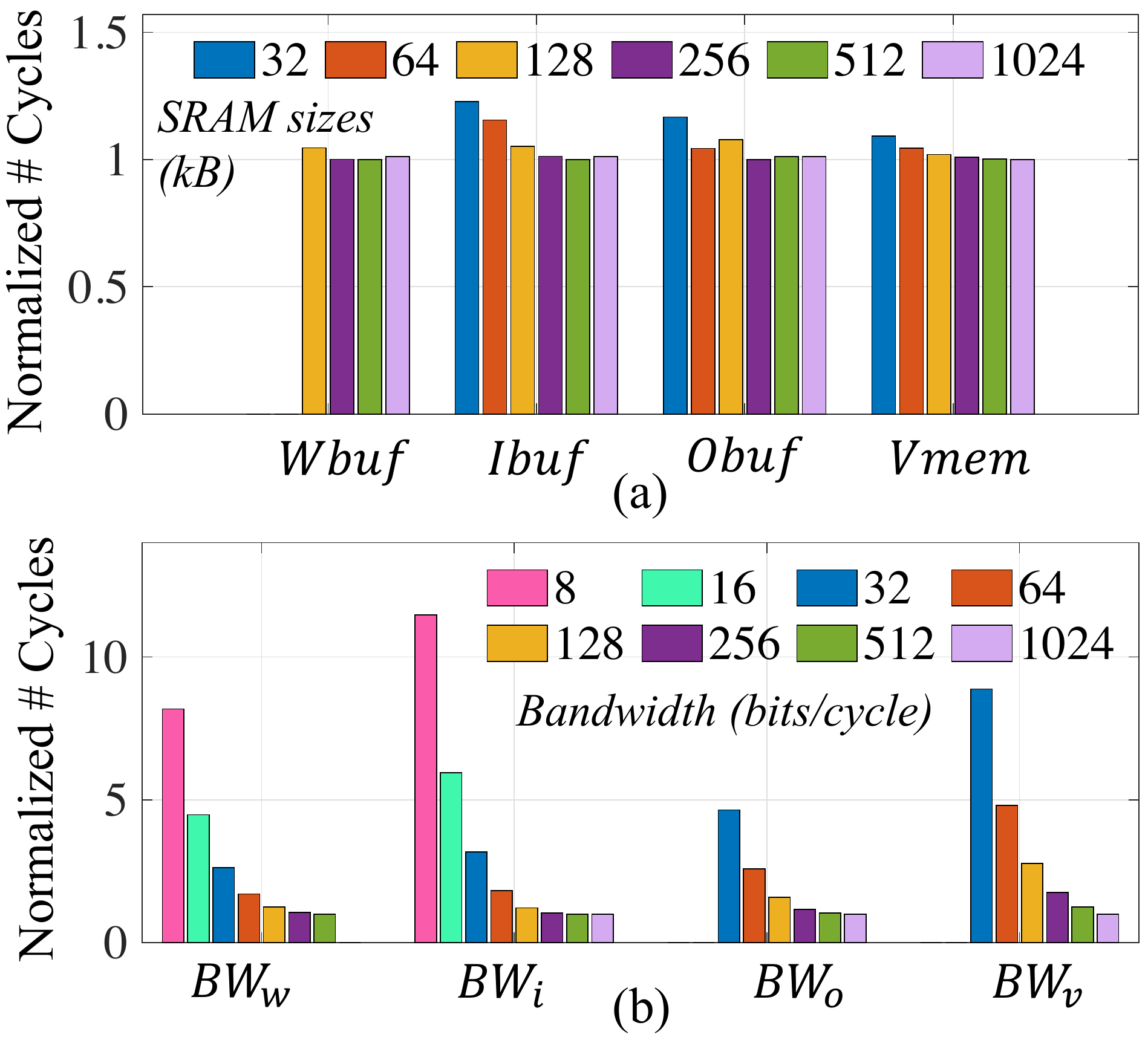}
	\vspace{-0mm}
	\caption{Normalized cycle counts of ResNet-50 inference on 64$\times$64 array with (a) various SRAM sizes and (b) various off-chip bandwidth parameters. (a missing bar indicates a hardware point that is not valid due to minimum/maximum size constraint on the respective hardware parameter).}
	\label{fig:SenPlot}
\end{figure}

\noindent
{\bf \textit{Sensitivity analysis:}} Finally, we perform sensitivity analysis to determine how each of the SRAM sizes and off-chip bandwidth parameters impacts the performance. Fig.~\ref{fig:SenPlot} shows the cycle counts of ResNet-50 inference for 64$\times$64 array (normalized to the cycle counts of the optimal hardware point obtained from Table~\ref{tbl:OptDist}). In (a), the results are shown by varying one SRAM size parameter at a time while keeping the remaining seven parameters same as the optimal configuration point (the small amount of non-monotonicity in the bars of (a) can be attributed to the tiling algorithm). Similar results are shown in (b) by varying one off-chip bandwidth parameter at a time. It is evident from (a) that the cycle counts have small sensitivity to the SRAM size parameters. For example, the smallest $Ibuf$ increases the cycle counts 1.23$\times$ from the optimal while the performance saturates quickly with an increment in $Ibuf$. The bars in (b) demonstrate that the performance is highly sensitive to the off-chip bandwidth parameters. For example, the smallest $BW_i$ shows 11.4$\times$ increment in cycle counts while the saturation starts from 256 bits/cycle bandwidth.

The Conv layers in the SA offer ample opportunity for data reuse, hence, changing one buffer size from the optimal configuration point does not impact the performance of SA significantly. Besides, since there is nearly zero data reuse opportunity in the non-Conv layers, the size of VMem minimally affects the DRAM access count and, thus, has small impact on the performance of the SIMD array. In contrast, since the bandwidth parameters dictate the costly data communication time with the off-chip DRAM, the performance is largely affected by these parameters.

\section{Conclusion}
\label{sec:Conclc}

\noindent
A comprehensive simulation framework covering convolution and a diverse set of non-convolution operations for DNN inference and training on ASIC accelerator platforms has been proposed. The detailed performance statistics provided by the simulator have been integrated with a backend flow to obtain end-to-end energy, runtime, and power of executing a network. Performance analysis using ResNet-50 and ResNet-18 training workloads reveals that non-convolution operations constitute a significant fraction of network runtime and energy. Design space exploration has been performed to identify optimal and 
economic inference design solutions under specified on-chip SRAM 
and off-chip bandwidth budgets. Optimal distribution of hardware resources is shown to offer substantial performance gain over generic static hardware resource allocation. In addition, economic design solutions are demonstrated to save remarkable memory footprint over the optimal hardware point at the cost of a small penalty in performance.

\section*{Acknowledgments}

\noindent
This work is supported in part by AFRL under the DARPA RTML program (award FA8650-20-2-7009). The U. S. government (USG) is authorized to reproduce and distribute reprints for governmental purposes notwithstanding any copyright notation thereon. The views and conclusions contained herein are those of the authors and should not be interpreted as necessarily representing the official policies or endorsements, either expressed or implied, of AFRL, DARPA, or USG.

\appendix
\section{Appendix}
\label{sec:appendix}

\subsection{Data Access and Cycle Count Models for BN$_{back}$}
\label{sec:BNBack_Models}

\noindent
The data access and cycle count models of SimDIT for BN$_{back}$ layer using the steps in Algorithm~\ref{alg:BN_grad_algo} is outlined here.
We present the derivations for Part-2. Similar models are developed for Part-1. The computation of BN$_{back}$ proceeds along the four computation loops (i.e., $h, w, n, c$). We define four multipliers to express the iterations\footnote{{\scriptsize Similar to Tensor-add layer, we present the models for BN$_{back}$ using outer tiles only, and all tiles and computation iterations correspond to outer tiles in this appendix.}} along these loops as follows (the tiling template is shown in Fig.~\ref{fig:BNGradTile}):
\begin{equation}
m_{h} = \frac{H}{T_{h}} \; \; ; \; \; 
m_{w} = \frac{W}{T_{w}} \; \; ; \; \; 
m_{n} = \frac{N}{T_n} \; \; ; \; \; 
m_{c} = \frac{C}{T_{c}}
\label{eq:OM_BN}
\end{equation}

\noindent
\underline{\textit{Number of DRAM accesses:}} There are two types of tensors in BN$_{back}$: 4D and 1D. The volume of a tile for a 4D tensor, ${\cal V}^{4D}_{tile}$, is given by ($T_h \cdot T_w \cdot T_n \cdot T_c$) while the volume of a 1D tensor tile, ${\cal V}^{1D}_{tile}$, equals $T_c$. During the computation of Part-2, each tiles of $\pmb {\gamma}^l$ (one 1D tensor), ${\bf \widehat{X}}^l$, $\frac{\partial L}{\partial {\bf X}^{l+1}}$, and $\frac{\partial L}{\partial {\bf X}^l}$ (three 4D tensors) are loaded/stored from/to DRAM once. Therefore, the number of DRAM access for Part-2 to compute the entire output tensor of $\frac{\partial L}{\partial {\bf X}^l}$ is given by:
\begin{equation}
A_{D_{Part-2}} = \left({\cal V}^{1D}_{tile} + 3{\cal V}^{4D}_{tile} (m_h m_w m_n)\right) m_c b_{io}
\label{eq:DRAM_access_BN}
\end{equation}
Here, $b_{io}$ is the data bit-width for all input and output tensors\footnote{{\scriptsize While the implementation of SimDIT accounts for separate bit-widths for input and output data, we use a single bit-width for clarity of the presented equations for BN$_{back}$.}}.

\noindent
\underline{\textit{Number of SRAM accesses:}} To compute the term outside the parenthesis in~\eqref{eq:data_grad}, one mul and one div operations per element of an 1D tile is required resulting in $2{\cal V}^{1D}_{tile}$ number of arithmetic operations. To compute the term inside the parenthesis and combine it with the term outside the parenthesis, three mul and two sub operations per element of a 4D tile is required. This leads to $5{\cal V}^{4D}_{tile}$ number of arithmetic operations. The total number of op count to process entire Part-2 is, therefore, computed by~\eqref{eq:Opcount_BN}. For each arithmetic operation, three VMem accesses are required (i.e., read access for two operands and write access for one operand). Hence, the number of SRAM accesses for Part-2 is computed from the op count using~\eqref{eq:SRAM_access_BN}.
\begin{align}
\mbox{op count} =& \; \left(2{\cal V}^{1D}_{tile} + 5{\cal V}^{4D}_{tile} \cdot (m_h m_w m_n)\right) m_c
\label{eq:Opcount_BN} \\
A_{S_{Part-2}} =& \; \mbox{op count} \cdot (2 + 1) \cdot b_{io}
\label{eq:SRAM_access_BN}
\end{align}

\noindent
\underline{\textit{Number of computation cycles:}} The SIMD array performs $K$ arithmetic operations in parallel where each ALU requires $\lambda_{\omega}$ cycles to complete an arithmetic operation $\omega$. Hence, the number of cycles to compute an 1D tile and a 4D tile of output (Lines~\ref{algo1:comp_cons} and~\ref{algo1:comp_out_grad} of Algorithm~\ref{alg:BN_grad_algo}) are computed using Equations~\eqref{eq:Comp_tile1D_BN} and~\eqref{eq:Comp_tile4D_BN}, respectively, where, as in Tensor-add, the $c$ dimension is mapped along the ALUs of the SIMD array.
\begin{align}
{\cal C}^{1D}_{tile} =& \; \left \lceil \frac{T_c}{K} \right \rceil \cdot (\lambda_{mul} + \lambda_{div})
\label{eq:Comp_tile1D_BN} \\
{\cal C}^{4D}_{tile} =& \; (T_h T_w T_n) \cdot \left \lceil \frac{T_c}{K} \right \rceil \cdot (3\lambda_{mul} + 2\lambda_{sub})
\label{eq:Comp_tile4D_BN}
\end{align}

\noindent
The tiled computations are iterated across the four loops to compute the entire $\frac{\partial L}{\partial {\bf X}^l}$ tensor. The total number of computation cycles, including pipeline setup overhead, for Part-2 is:
\begin{equation*}
{\cal C}_{Part-2} = \left[\left({\cal C}^{4D}_{tile} + PSO_{SIMD} \right) (m_h m_w m_n) + {\cal C}^{1D}_{tile}\right] m_c
\label{eq:Comp_BN}
\end{equation*}

\noindent
\underline{\textit{Number of stall cycles:}} The SIMD array is stalled during the load and store operations of each tile. Thus, using the tiled volumes of data and off-chip bandwidth for VMem, the number of stall cycles to process Part-2 is computed by:
\ignore{
{\footnotesize
\begin{align}
{\cal S}^{BN_{back}} = &\left \lceil \frac{{\cal V}^{1D}_{tile} \cdot b_{io}}{BW_v}\right \rceil \cdot m_c \nonumber \\
&+  \left \lceil \frac{3{\cal V}^{4D}_{tile} \cdot b_{io}}{BW_v}\right \rceil \cdot (m_h \cdot m_w \cdot m_n \cdot m_c)
\label{eq:Stall_BN}
\end{align}
}
}
\begin{equation*}
{\cal S}_{Part-2} = \left(\left \lceil \frac{{\cal V}^{1D}_{tile} \cdot b_{io}}{BW_v}\right \rceil + \left \lceil \frac{3{\cal V}^{4D}_{tile} \cdot b_{io}}{BW_v}\right \rceil (m_h m_w m_n)\right) m_c
\label{eq:Stall_BN}
\end{equation*}


\bibliographystyle{IEEEtran}
\bibliography{main}


\end{document}